\documentclass{aa}
\usepackage{graphicx}
\usepackage{txfonts}
\usepackage{natbib}

\begin{document}
\title{Simulated SKA maps from Galactic 3D-emission models}
\author{X.~H.~Sun\inst{1,2}, \and W.~Reich\inst{2}} 
\offprints{Wolfgang Reich\\ \email{wreich@mpifr-bonn.mpg.de}}
\institute{National Astronomical Observatories, CAS, 
           Jia 20, Datun Road, Chaoyang District, Beijing 100012, China
      \and
           Max-Planck-Institut f\"{u}r Radioastronomie, Auf dem H\"{u}gel 69, 
           53121 Bonn, Germany}

\date{Received ; accepted }

\abstract 
{
 Planning of the Square Kilometre Array (SKA) requires simulations of the 
 expected sky emission at arcsec angular resolution to 
 evaluate its scientific potential, to constrain its technical realization 
 in the best possible way and to guide the observing strategy.
}
{
 We simulate high-resolution total intensity, polarization and rotation 
 measure (RM) maps of selected fields based on our recent global 3D-model of 
 Galactic emission.
}
{
 Simulations of diffuse Galactic emission are conducted using the 
\textsc{hammurabi} code modified for arcsec angular resolution patches towards 
various Galactic directions. The random magnetic field components are set to 
follow a Kolmogorov-like power law spectrum.
}
{
 We present maps for various Galactic longitudes and latitudes at 
 1.4~GHz, which is the frequency where deep SKA surveys are proposed. The maps 
 are about $1\fdg5$ in size and have an angular resolution of about $1\farcs6$.
 We analyse the maps in terms of their probability density functions (PDFs) 
 and structure functions. Total intensity emission is more smooth in the plane 
 than at high latitudes due to the different contributions from the regular and
 random magnetic field. The high latitude fields show more extended polarized 
 emission and RM structures than those in the plane, where patchy emission 
 structures on very small scales dominate. The RM PDFs in the plane are close 
 to Gaussians, but clearly deviate from that at high latitudes. The RM 
 structure functions show smaller amplitudes and steeper slopes towards high 
 latitudes. These results emerge from the fact that much more turbulent cells 
 are passed through by the line-of-sights in the plane. Although the simulated 
 random magnetic field components distribute in 3D, the magnetic field spectrum
 extracted from the structure functions of RMs conforms to 2D in the plane and 
 approaches 3D at high latitudes. This is partly related to the outer scale of 
 the turbulent magnetic field, but mainly to the different lengths of the 
 line-of-sights.
}
{
 The significant scatter of the simulated RM distributions agrees with the 
 large scatter of observed RMs of pulsars and extragalactic sources in the 
 Galactic plane and also at high Galactic latitudes. A very dense grid of RMs 
 from extragalactic sources is required to trace and separate Galactic 
 RM foreground fluctuations. Even at high latitudes total intensity and 
 polarized emission is highly structured, which will contaminate sensitive 
 high-resolution extragalactic observations with the SKA and have to be 
 separated in an appropriate way. 
}


\titlerunning{High resolution simulations}
\authorrunning{X.~H.~Sun \& W.~Reich}

\maketitle

\section{Introduction}
Intensive radio emission from the Galaxy originates from ionized thermal gas, 
cosmic rays and magnetic fields. All-sky surveys yield an overview of the 
Galactic structures. However, to reveal distributions of the Galactic emission 
components, modelling is required because of our position inside the Galaxy. 
Galactic emission is a severe foreground contamination for observations of 
fluctuations of the cosmic microwave background radiation and of all kinds of 
distant, faint extragalactic objects, which will be investigated by the Square 
Kilometre Array (SKA) with unprecedented sensitivity. Therefore it is crucial 
to properly account for Galactic emission in all directions of the sky. 
Unfortunately, most of the properties of the aforementioned constituents of the
 Galaxy from large scales to small scales are insufficiently constrained up to 
now.

The large-scale Galactic magnetic field is of particular interest. It has been 
intensively studied by using rotation measures (RMs) of pulsars 
\citep[e.g.][]{hml+06,njkk08} and extragalactic sources (EGSs) in the Galactic 
plane \citep[e.g.][]{btj03,bhg07}. Pulsar dispersion measures (DMs) are used to
 find their distances based on a thermal gas distribution model. Recent pulsar 
results indicate a number of magnetic field reversals within the Galaxy 
\citep[e.g.][]{hml+06,njkk08}, which is difficult to understand in terms of a 
simple axi-symmetric or bi-symmetric magnetic field configuration. A more dense
 RM grid is needed to derive a clear large-scale pattern, because complications 
arise from local magnetic field disturbances or excessive thermal gas 
contributions to the observed pulsar DMs and RMs \citep{mwkj03}. Such effects 
must be separated to derive a reliable large-scale magnetic field configuration.
 Moreover, to decompose the observed RM information into a magnetic field 
contribution, a detailed knowledge of the thermal gas distribution is required. 
The currently widely accepted distribution of Galactic diffuse thermal gas, the
 NE2001 model, as well as its filling factor modelled from pulsar and 
H$\alpha$ observations was presented by \citet{cl02} and \citet{bmm06}, 
respectively. Galactic magnetic field models based on pulsar RM observations 
have to agree with the observed RMs of extragalactic sources shining through 
the entire Galaxy. To establish a firm picture of the large-scale magnetic 
field needs to account for all the effects above.
 
Recently, \citet{srwe08} have derived a Galactic 3D-emission model which  
reproduces all-sky surveys over a wide frequency range and also the all-sky RM 
distribution of EGSs. The simulated maps have an angular resolution of about 
15$\arcmin$, which is suitable to compare them with all-sky maps having 
lower angular resolution. Based on these global results for the Galaxy, 
simulated arcsec angular resolution maps are the next step in order to get an 
idea what might be observed with the SKA and other future high-resolution 
radio telescopes. It is important to understand the origin of the observed 
structures at high angular resolution to estimate the contamination of 
extragalactic observations by Galactic small-scale structures as the foreground.

One of the key science projects for the SKA is ``Cosmic Magnetism", 
where RMs of about $2\times10^7$ polarized EGSs are proposed to be observed in 
a region of $10^4$~deg$^2$ \citep{bg04}. The survey is proposed for 
1.4~GHz with a resolution of about $1\arcsec$. Simulations of the expected 
number density of faint sources and their polarization properties from SKA 
observations were made by e.g. \citet{wmj+08}. A feasibility study on the 
decomposition of the source RMs from the Galactic foreground RMs and its 
removal need high-resolution Galactic RM maps. This is the basic motivation for
 extending our simulations to selected Galactic patches with arcsec angular 
resolution.  
   
Fluctuations of the components of the interstellar medium (ISM) dominate 
the emission structures at high angular resolution. The thermal electron 
density distribution has been found to follow a Kolmogorov-like power law from 
about $10^8$~m to about $10^{18}$~m \citep{ars95}. The fluctuations of the 
Galactic magnetic field are mainly studied based on the analysis of the 
structure function of RMs, where an additional effort is needed to decouple the 
magnetic field and the electron density fluctuations \citep{ms96}. 
The outer scale of turbulence was reported to be in the range of about 
100~pc \citep{ars95,hbgm08} down to a few pc \citep{ms96,hbgm08}. 

We summarize our global Galactic 3D-modelling in Sect.~2. The high angular 
resolution simulation method and the treatment of small-scale magnetic fields 
are described in Sect.~3. Results and analysis of the simulated RM maps are 
presented in Sect.~4, including a discussion of the consequences of the 
simulated RM fluctuations. In Sect.~5 we discuss the results of the simulated 
total intensity and polarization maps. Some general conclusions are given in 
Sect.~6. 

\section{A new Galactic 3D-emission model}

\citet{srwe08} presented a new Galactic 3D-emission model, which is in
agreement with a wide range of observations:

\begin{itemize}
\item Galactic synchrotron radiation being best represented by the 408~MHz
      all-sky survey \citep{hss+82} in total intensity and by the WMAP 22.8~GHz
      all-sky survey in polarized intensity \citep{phk+07}.

\item Optically thin thermal emission by the WMAP 22.8~GHz free-free emission
      template \citep{hin07}, and optically thick thermal emission causing 
      absorption, which reflects in a spectral flattening at low frequencies 
      along the Galactic plane.

\item The all-sky 1420~MHz polarization survey combined from northern 
      \citep{wlrw06} and southern sky data \citep{trr08} constraining
      depolarization properties of the interstellar medium.

\item RM data from EGSs along the Galactic plane taken from the Canadian (CGPS) 
      and the Southern Galactic Plane Survey (SGPS) \citep{btj03,bhg07}, and 
      EGSs out of the plane from \citet{hmq99}, including preliminary RMs from 
      the new Effelsberg L-band survey (Han et al., in prep.), and RMs from the
      CGPS high latitude extension (Brown et al., in prep.).
\end{itemize}

The  Galactic 3D-model is based on the 3D-distribution of thermal electrons 
in the Galaxy (NE2001 model) derived from pulsar DMs by \citet{cl02}. To 
successfully model the observed thermal emission for the optically thin case 
and absorption effects from the optically thick gas, the filling factor of 
thermal gas needs to be taken into account, whose spatial variation was derived
 by \citet{bmm06}. The regular Galactic disk magnetic field model was 
constrained by extragalactic RM data of \citet{btj03,bhg07}. Axi-symmetric 
(ASS) and bi-symmetric (BSS) fields with one field reversal inside the solar 
circle agree with extragalactic RM data in the plane, while the BSS field 
configuration fails to represent the RM gradient with Galactic latitude as 
measured from the high-latitude extension of the CGPS by Brown et al. 
(in prep.). The 408~MHz survey \citep{hss+82} and the WMAP 22.8~GHz linear 
polarization survey \citep{phk+07} were used to constrain the random magnetic 
field component, the cosmic-ray electron distribution, and the known local 
synchrotron excess. Finally, the depolarization effects of the 1420~MHz 
polarization survey \citep{wlrw06,trr08} were modelled, which requires to 
modify the thermal filling factor by adding an extra coupling term of the 
electron density and the random magnetic field. 

The quantitative model parameters of the three constituents of the Galaxy are 
those of the NE2001 thermal electron density model, an ASS magnetic field 
configuration with a local strength of 2~$\mu$G, asymmetric toroidal halo 
fields with respect to the plane with a maximum strength of 10~$\mu$G, 
isotropic and homogeneous random magnetic fields with a strength of 3~$\mu$G, 
and a cosmic-ray electron distribution truncated at 1~kpc above and below the 
plane. The details are given by \citet{srwe08}.

Although this Galactic 3D-model is in agreement with the observed all-sky maps,
some of its parameters seem unrealistic. To fit the observed high-latitude RMs,
a strong regular halo magnetic field is required, because the scale height of 
the thermal electrons of the NE2001 model is just about 1~kpc. The strong halo 
field in turn requires a cosmic-ray electron distribution limit of about 1~kpc 
to avoid excessive high-latitude synchrotron emission. These parameters are 
unrealistic for both the halo field strength \citep{ms08} and the distribution 
of cosmic-ray electrons. \citet{srwe08} note that by increasing the scale 
height of the thermal electron density to about 2~kpc the halo magnetic field 
strength reduces to about 2~$\mu$G and the truncation of the cosmic-ray 
distribution turns obsolete. Recently, \citet{gmcm08} critically reanalyzed the
 thermal electron scale height and quote a revised value of 1.8~kpc, very close
 to the proposed 2~kpc by \citet{srwe08}. This makes the Galactic properties 
similar to those observed for nearby galaxies. It should be noted that the 
simulated maps are not affected by these modifications of parameters.

\section{High-resolution simulations}

\subsection{The general simulation concept}

The simulations presented in this paper make use of the \textsc{hammurabi} code 
\citep{wae05,wjr+09}, where the sky is pixelized according to the HEALPix 
scheme \citep{ghb+05}. Each pixel contains information from a cone shaped 
volume centered at the observer. Each surface element is quadrilateral and 
covers the same area. The number of pixels $N_{\rm pix}$ of a sphere 
are calculated by $N_{\rm pix}=12N^2_{\rm SIDE}$, where $N_{\rm SIDE}$ is 
an integer power of 2. Therefore d$\Theta$, the linear size or the angular 
resolution, can be calculated as 

\begin{equation}\label{resolution}
{\rm d}\Theta=\sqrt{\frac{3}{\pi}}\frac{3600\arcmin}{N_{\rm SIDE}}.
\end{equation}        

For each pixel of the map, the Stokes parameters ($I$, $U$ and $Q$) and the 
RM are integrated in the radial direction from the observer up to a maximal 
distance $r_{\rm max}$ \citep{srwe08}. The modelled emission distributions 
in the Galaxy are represented in cylindrical coordinates with the origin at the 
Galactic center. The distance to the Sun is taken as 8.5~kpc. Details of the 
regular magnetic field component, the cosmic-ray density and the thermal gas 
distributions in the disk and halo of the Galaxy are described by 
\citet{srwe08}. The integration process samples these distributions and 
inherently accounts for depolarization. To accurately calculate the integral, 
each pixel is separated into many volume units with equal radial interval 
$\Delta r$. Thus the size of a unit can be approximated as 
$r{\rm d}\Theta\times r{\rm d}\Theta \times \Delta r$ with $r$ being the 
distance of the volume unit to the observer. This process is repeated for all 
pixels, which are finally combined to an all-sky map.

\subsection{Zoom-in for specified patches}

According to Eq.~(\ref{resolution}) the possible resolutions depend on 
$N_{\rm SIDE}$. To achieve arcsec resolution and to optimize the amount of 
computation time, we use $N_{\rm SIDE}$ of 131072, which corresponds to a 
resolution of about $1\farcs6$. The pixel number for an all-sky map then 
amounts to about $2\times10^{11}$, which is far beyond any computer memory 
capacity today. Therefore high-resolution simulations are just possible for a 
small patch of sky. Thus we adapted the \textsc{hammurabi} code to be able to 
simulate parts of the sky.

\begin{figure}[!htbp]
\includegraphics[angle=-90,width=8.5cm]{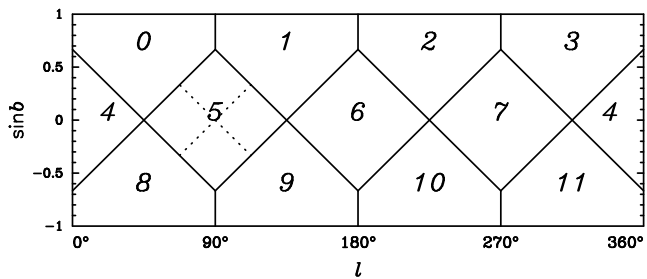}
\caption{The HEALPix pixel scheme \citep{ghb+05} for Galactic coordinates used 
in this paper.}
\label{healpix}
\end{figure}

In the \textsc{hammurabi} code the NEST ordering \citep{ghb+05} is used, which 
facilitates the zoom-in of specified patches as it is explained below.  For 
$N_{\rm SIDE} = 1$ the entire sky is split into 12 patches as shown in 
Fig.~\ref{healpix}. They are numbered from 0 to 11. For $N_{\rm SIDE}=2$, each 
pixel is further split into four child pixels, also shown in Fig.~\ref{healpix}
 for patch ``5''. This patch is centered at longitude of $90\degr$ in the plane 
with a diagonal size of about $90\degr \times 90\degr$. Subsequently for 
$N_{\rm SIDE}$ increasing by 2 each pixel is further split into four pixels. 
In this scheme one can determine a target patch and make proper splits 
depending on the resolution. To reach a smaller patch size and higher angular 
resolution a much larger value for $N_{\rm SIDE}$ is required.  
 
\subsection{Realization of the random magnetic field component}
\label{b_realization}

It is generally accepted that the power spectrum of the random magnetic field 
component should follow a power law as $P(k)\propto k^\alpha$. Here $k=2\pi/l$ 
is the wave vector and $l$ is the spatial scale. For a Kolmogorov-like 
turbulence spectrum the spectral index is $\alpha=-11/3$, which is used in all 
present simulations. Note that independent fluctuations of the electron 
density are not included in the simulations. Howver, fluctuations of the 
electron density are also expected to follow a Kolmogorov-like spectrum 
\citep{ars95}. The coupling of electron density and random magnetic fields we 
introduced in our all-sky simulations to model the observed depolarization 
\citep{srwe08} results in Kolmogorov-like fluctuations for both components.
 
Random magnetic field components realized in the simulations presented by 
\citet{srwe08} are generated {\it in situ} for every volume unit by Gaussian 
random numbers\footnote{Throughout all the simulations we use the random 
number generator from the GNU Scientific Library. The algorithm is MT19937, 
which has a Mersenne prime period of $2^{19937}-1$.}. To obtain the random 
magnetic field components following a given power-law spectrum, an inverse 
Fourier transformation must be performed. All available fast Fourier 
transformation (FFT) algorithms are based on a Cartesian coordinate system. We 
therefore set-up a cubic box with a size of $L^3$ and divide it into $N^3$ 
small cubes with a common volume of $\Delta^3$. Within the box we construct a 
Cartesian coordinate system and perform the FFT to derive the random field 
strength and direction in each small cube. Here $L$ corresponds to the outer 
scale of the turbulence and $\Delta$ corresponds to the inner scale.  

An important constraint for $\Delta$ is that $\Delta\leq r{\rm d}\Theta$. This 
ensures that several neighbouring line-of-sights do not pass the same small 
cubes. Otherwise spurious stripes are seen in the maps. For $r$ of about 1~kpc 
and ${\rm d}\Theta$ of $1\farcs6$ the size of a small cube should not be larger
 than about 0.01~pc. If our Galaxy is treated as one box, meaning $L=40$~kpc, 
the total number of cubes is about $6.4\times10^{19}$. Again there is no way in
 processing this huge amount of cubes by current computing facilities.    

In our simulation, we devise a scheme to generate the random magnetic field
components in an adequate way. We first conduct a FFT to create a magnetic 
field distribution in a box. We then rotate this box to acquire new boxes,
which are glued together to fill the entire Galaxy. The box rotation angles are
 integer multiples of $\pi/2$, which avoids cutting of boxes. Through this 
scheme the turbulence status of the Galaxy is realized. This process is 
detailed in Appendix~\ref{b_fft}. 

Actually the size of the box cannot be too small, otherwise there are too many 
boxes to handle. On the other hand $L$ cannot be too large in order to get a 
reasonable number of cubes of about 0.01~pc in size. As a compromise, we take 
a size of 10~pc for the box and of 0.0125~pc for the cubes. Although 
0.0125~pc is much larger than the inner scale of the electron density 
fluctuations, which are of the order of $10^8$~cm \citep[e.g. ][]{ars95}, it is 
sufficient for our study in regard to the current angular resolution. In total 
there are about $5\times10^8$ cubes, requiring a memory of about 12~giga-byte 
for the random magnetic field components when stored as double precision 
numbers. It is thus obvious that the number of cubes is a compromise 
between high resolution, which requires small cube sizes, and hence a very 
large number of cubes and the available computer memory, which is thus able to 
handle a limited number of cubes.

\begin{figure*}[!htbp]
\centering
\includegraphics[angle=-90,width=18cm]{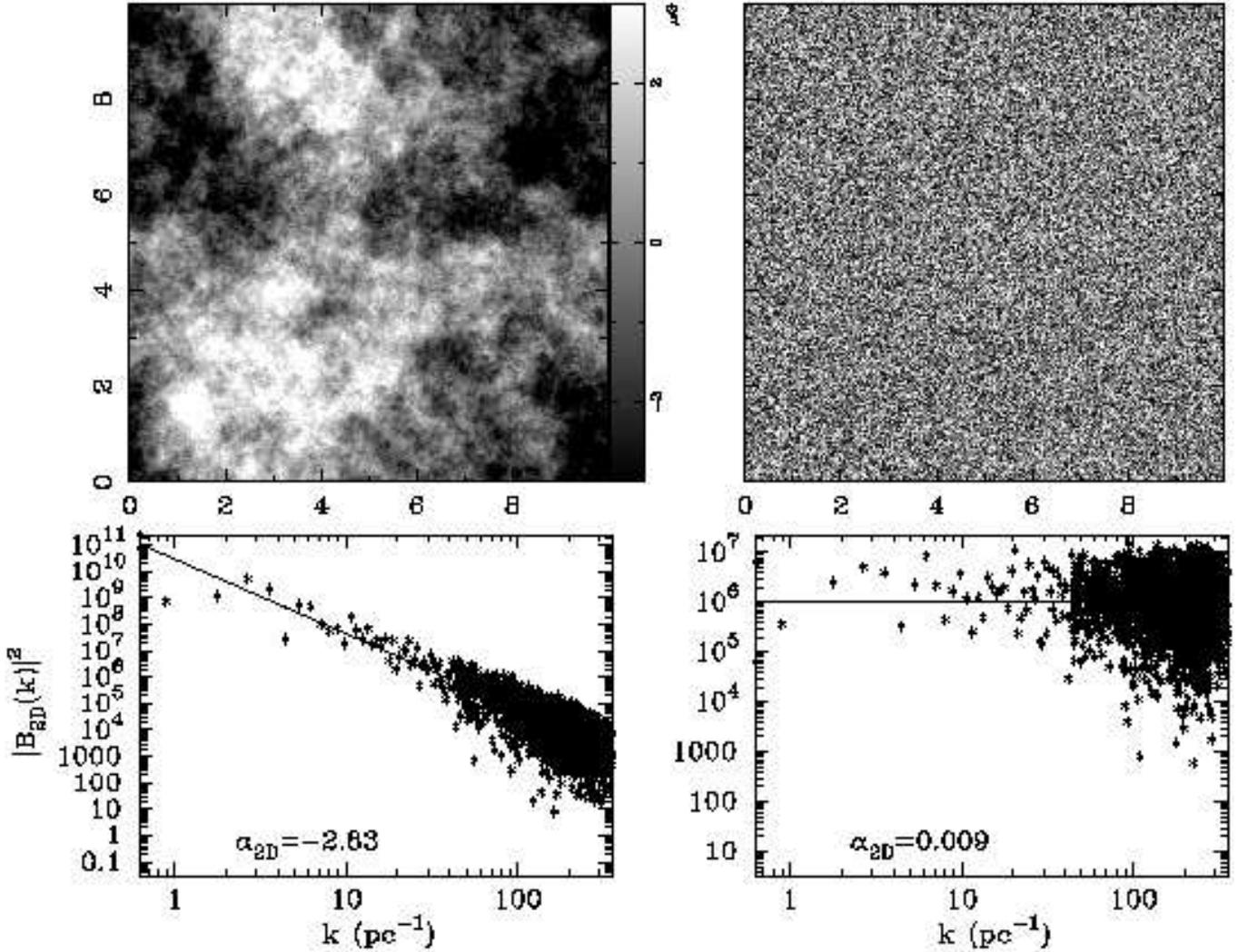}
\caption{2D random magnetic field distribution in a slice parallel to one 
surface of a 10~pc box. The upper left panel shows magnetic field components 
following a power law index of $-11/3$, while the upper right panel shows 
random field components realized by random Gaussian numbers corresponding to a 
spectral index of zero. Their corresponding power spectra are shown in the 
lower panels.}
\label{bslice}
\end{figure*}

In Fig.~\ref{bslice}, we show the 2D-distribution of random magnetic field 
components for a slice parallel to one surface of a 10~pc box. For 
comparison, we show the random magnetic fields generated {\it in situ} in each 
grid of the same plane as random Gaussian number, which is similar to that used
 by \citet{srwe08}. The power spectra of the magnetic fields are also shown in 
Fig.~\ref{bslice}. It is apparent that the distribution of random magnetic 
fields following a power-law spectrum shows coherent structures of various 
extent, whereas the Gaussian random magnetic field distribution is featureless.
 The spectral index of $\alpha_{\rm 2D}=-2.83\approx\alpha+1$ for Kolomogorov 
magnetic field distributions and $\alpha_{\rm 2D}=0$ for Gaussian field 
distributions agree with the expectations and prove that the random magnetic 
field components are properly distributed in the box. 

\section{Modelling results on rotation measures}

Except for the random magnetic fields the simulations presented below use the 
same Galactic 3D-emission models as derived by \citet{srwe08}, where the 
``ASS+RING" model for the disk magnetic field is selected here. The random 
magnetic fields are introduced as described above and follow a Kolmogorov-like 
spectrum. The strength of the random fields is scaled to 3~$\mu$G which is the 
same as that in \citet{srwe08}. All simulations are made for 1.4~GHz 
observations, as this is the proposed SKA RM survey frequency. As mentioned 
above, the simulated patch size is limited by available computer memory, so 
that just relatively small patches of about $1\fdg5$ in size are simulated at a
 resolution of $1\farcs6$. This is compatible with the expected SKA angular 
resolution. 

Simulations are presented for a handful of patches only, because each run takes
 very long computation time. We select the first patch around 
$(l,\,\,b)\,=\,(135\degr,\,\,+40\degr)$, which is in the direction of the 
DRAO {\it Planck} Deep Fields \citep{tsj+07}, where soon high-resolution 
continuum and polarization data will be available. This patch is named MN 
standing for Medium-latitude-North. More patches with roughly the same 
longitude, but with different latitudes, are further selected. These are 
labeled HN (High-latitude-North), PNAC1 (Plane-North-Anti-Centre 1), PSAC 
(Plane-South-Anti-Centre), MS (Medium-latitude-South) and HS 
(High-latitude-South). To compare with PNAC1, three additional patches in the 
plane, PNAC2 (Plane-North-Anti-Centre 2), PNC1 and PNC2 
(Plane-North-Centre 1 and 2) are included. The centre coordinates of these 
patches are listed in Table~\ref{patch}. Note that we cannot set the patch 
coordinates to exactly the same longitudes or latitudes because of HEALPix 
constraints \citep{ghb+05}. In Table~\ref{patch} also the maximal integral 
lengths or the length of line-of-sights ($r_{\rm max}$) and the integral steps 
($\Delta r$) are listed. The integral length is determined by taking into 
account the cosmic-ray scale height of about 1~kpc and the disk cutoff radius 
of 17.5~kpc for the electron density in the NE2001 model \citep{cl02}. 
Integrating further out results in negligible contributions. For the patches 
with large integral length, the step size is larger than for those with small 
length to limit the computation time. The selected patches allow to study 
variations of emission and RM structures related to turbulent magnetic field 
components versus Galactic longitudes and latitudes. Variations of the 
structure functions for RMs from EGSs as a function of Galactic direction have 
already been noted by \citet{sh04} for angular scales of several degrees. 

\begin{table}[!htbp]
\caption{Simulated patches.}
\label{patch}
\begin{tabular}{crrcc}\hline\hline
name &$l$ ($\degr$) &$b$ ($\degr$)&$r_{\rm max}$ (kpc)&$\Delta r$ (pc)\\ \hline
HN    &  138.33 &   70.16 &  3            &   1.0  \\ 
MN    &  136.48 &   44.20 &  3.5          &   1.0  \\ 
PNAC1 &  130.07 &    1.19 & 10            &   2.5  \\
PSAC  &  130.07 & $-$1.19 & 10            &   2.5  \\
MS    &  136.48 &$-$44.20 &  3.5          &   1.0  \\
HS    &  138.33 &$-$70.16 &  3            &   1.0  \\
PNC1  &   59.77 &    1.19 & 30            &   3.8  \\
PNC2  &  300.23 &    1.19 & 30            &   3.8  \\
PNAC2 &  229.92 &    1.19 & 10            &   2.5  \\
\hline
\end{tabular}
\end{table}

All simulations were made by using the computer cluster of the 
Max-Planck-Institut f\"ur Radioastronomie operated by the VLBI group. 

\subsection{Overview of the results}

As described by \citet{srwe08} the \textsc{hammurabi} simulations result in 
Stokes $I$, $U$, and $Q$ maps at the specified frequency as well as the RM 
maps. Subsequently, the polarization angle ($PA$) and polarized intensity 
($PI$) maps are calculated according to $PA=\frac{1}{2}{\rm atan}\frac{U}{Q}$ 
and $PI=\sqrt{U^2+Q^2}$.

\begin{figure*}[!htbp]
\centering
\includegraphics[width=18cm]{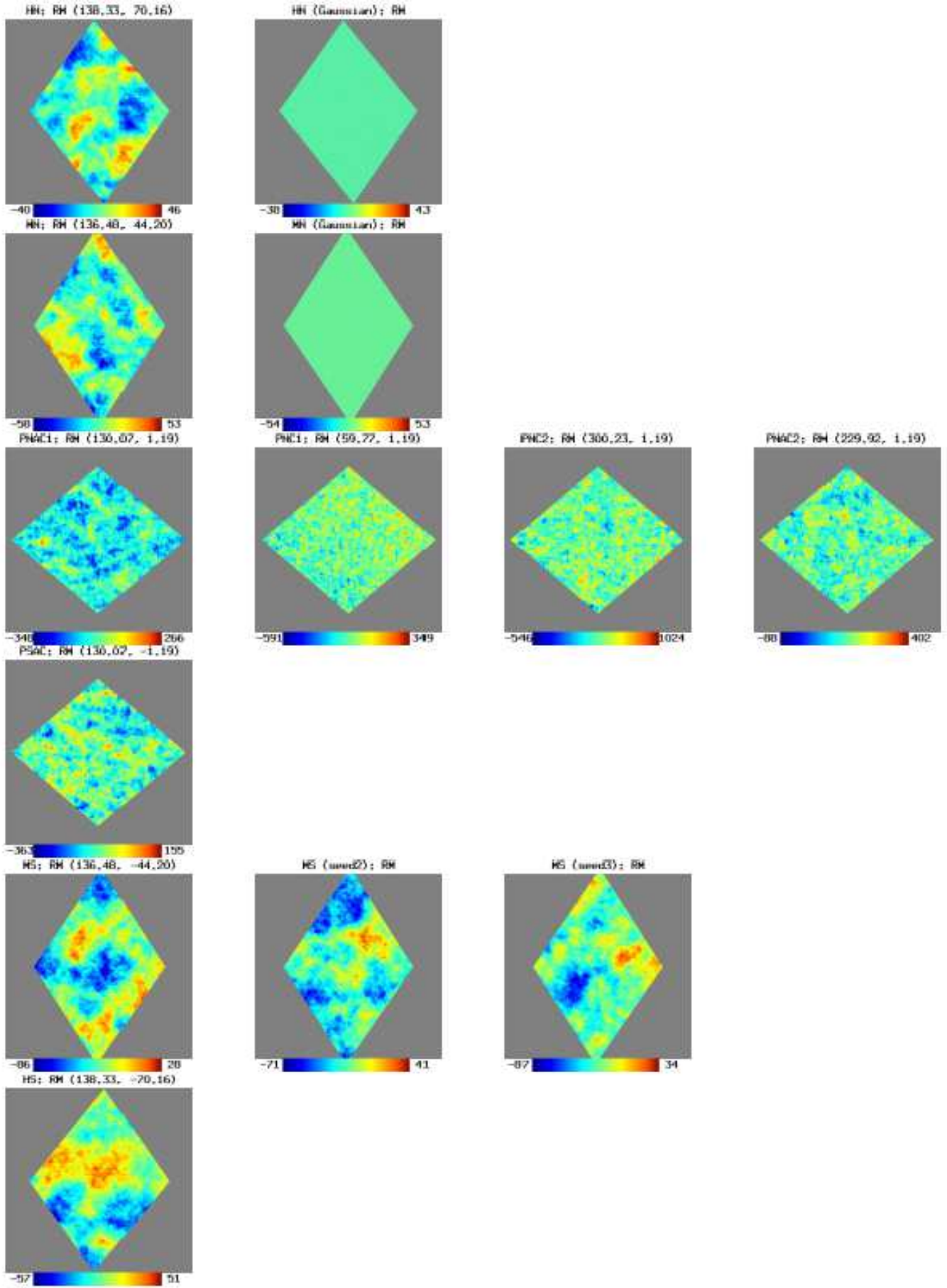}
\caption{RM maps for all simulated patches ordered from high to low latitudes 
(top to bottom) in gnonomic projection. Their designation and Galactic 
coordinates are quoted on top of each individual panel (also listed in Table~1).
 Different simulations for the same patch and for patches at the same latitude,
 but different longitudes, are shown in one row (see text for details).  Each 
field has about $3000 \times 3000$ pixel and a size of about 
$1\fdg5\times1\fdg5$. The minimum and maximum of each map is shown together 
with the wedge below each panel. }
\label{rmmaps}
\end{figure*}

In this paper, we will not discuss the separation process of total and 
polarized intensities of EGSs from the foreground, which requires adding 
polarized EGSs and noise to the simulated maps. The simulated RMs represent the
 contribution of the whole Galaxy to the RM from an EGS in case the source size
 does not exceed the pixel size of the simulated map. Thus, the RM maps are 
directly related to observations. Below we focus on the analysis of RM maps, 
which are directly connected to the planned SKA RM survey. All simulated RM 
maps are shown in Fig.~\ref{rmmaps}. The other maps are shown in 
Figs.~\ref{imaps}, \ref{umaps}, \ref{qmaps}, \ref{pimaps}, and \ref{pamaps} 
in Appendix~\ref{othermaps}.   

The RM maps clearly show structures ranging from small to large scales. 
The variation of the structures versus latitudes is also evident. Below some 
statistical methods are applied to analyse these structures.    

\subsection{Statistical description of the simulations}

The rotation measure, RM, is defined as
\begin{equation}\label{rmeq}
{\rm RM}=K\int_0^{r_{\rm max}} n_e (B_\parallel+b_\parallel) {\rm d}r,
\end{equation}
where $K$ is a constant, $n_e$ is the thermal electron density, $B_\parallel$ 
and $b_\parallel$ are the regular and random magnetic field components 
projected along the line-of-sight, respectively, and $r_{\rm max}$ is the 
maximal distance along the line-of-sight from the observer. One major question 
is to what extent the properties of $B$ and $b$ can be recovered from the RM 
maps.  

\subsubsection{The probability distribution function}

We use the probability distribution function (PDF) to describe the structure 
properties of a map as discussed by \citet{mrf+07}. For the RM distribution 
$\rm{PDF}~({\rm RM})\times\Delta {\rm RM}$ quantifies the relative number of 
pixels falling between ${\rm RM}$ and ${\rm RM}+\Delta {\rm RM}$. A Gaussian 
shape of the PDF indicates that no extended features are present in the map. 
To measure the deviation from a Gaussian distribution, two more parameters, 
skewness and kurtosis, are used \citep[e.g.][]{klb07},
\begin{equation}
\begin{array}{rcl}
{\rm skewness}\,&=&\displaystyle{\frac{1}{N}\sum_{i=1}^{N}
\left(\frac{{\rm RM}_i-\overline{\rm RM}}{\sigma_{\rm RM}}\right)^3}\\[5mm]
{\rm kurtosis } \,&=&\displaystyle{\frac{1}{N}\sum_{i=1}^{N}
\left(\frac{{\rm RM}_i-\overline{\rm RM}}{\sigma_{\rm RM}}\right)^4-3},
\end{array}
\end{equation}
where $\overline{\rm RM}$ and $\sigma_{\rm RM}$ are the average and 
standard variance of the RM maps. They are third and fourth order statistical
 moments, respectively. Large absolute values of skewness or kurtosis indicate 
that either the tail or the centre of the Gaussian is deformed. Physically 
these parameters are closely related to turbulence properties \citep{klb07}. 

The RM PDFs for all simulated patched are shown in Figs.~\ref{pdfsf_hnmn}, 
\ref{pdfsf_ms}, \ref{pdfsf_pn}, and \ref{pdfsf_hmp} (upper panels). Their 
average values, together with variance, skewness and kurtosis are listed in 
Table~\ref{statistic_par}.     

\begin{table}[!htbp]
\caption{Statistical parameters for the simulated RM patches. The patch 
designations are listed in Table~1. Note that for the patches HN, MN and MS 
more realizations were made with details given in the text.}
\label{statistic_par}
\centering
\begin{tabular}{lrrrr}\hline\hline
name   & average   &  variance  &  skewness & kurtosis\\
      & (rad m$^{-2}$) & (rad m$^{-2}$) & & \\\hline
HN            &     2.50 &  12.73&   0.0857& $-$0.2555 \\
HN (Gaussian) &     2.14 &   8.11&$-$0.0007&    0.0008 \\
MN            &  $-$1.90 &  15.12&   0.0347& $-$0.2287 \\
MN (Gaussian) &     1.41 &  10.93&$-$0.0007&    0.0009 \\
PNAC1         & $-$85.82 &  63.40&   0.0654&    0.2124 \\
PSAC          &$-$103.13 &  60.86&$-$0.0482& $-$0.0274 \\
MS (seed1)    & $-$27.58 &  18.81&$-$0.1464& $-$0.6365 \\
MS (seed2)    & $-$19.78 &  16.74&   0.1338& $-$0.2968 \\
MS (seed3)    & $-$24.86 &  16.35&$-$0.1847&    0.3896 \\
HS            &     1.66 &  17.15&$-$0.1241& $-$0.6353 \\
PNC1          & $-$89.03 &  93.65&$-$0.0018&    0.0646 \\
PNC2          &   286.53 & 155.71&$-$0.0045&    0.0528 \\
PNAC2         &   159.80 &  50.80&   0.0217& $-$0.0429 \\
\hline
\end{tabular}
\end{table}

\subsubsection{The structure function}

For turbulence studies, it is customary to investigate the power spectrum of 
the magnetic field, which can be retrieved from the power spectrum, 
autocorrelation function, or structure function of RMs  
\citep[e.g.][]{scs84, ms96,ev03}. Observationally always uneven sampled RMs of 
EGSs are obtained. In this case the structure function is suited best 
\citep{sh04}. Since RM maps are available as simulation results, the 
autocorrelation function is also investigated here as a crosscheck for the 
structure function results. The structure function ($D_{\rm RM}(\delta\theta)$) 
is defined as 
\begin{equation}
D_{\rm RM}(\delta \theta)=\frac{[{\rm RM}(\vec{\theta})-
{\rm }RM(\vec{\theta+\delta \theta})]^2}{N_{\rm pairs}},
\end{equation}    
where $\delta \theta$ is the angular separation of two pixels and 
$N_{\rm pairs}$ is the number of RM pairs with the same angular separation.  

\begin{figure}[!htbp]
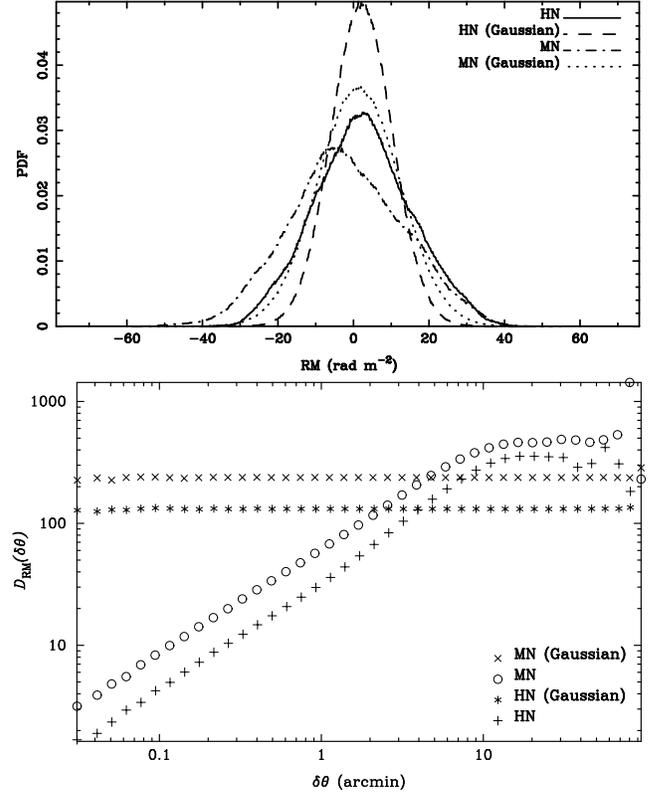

\centering
\includegraphics[width=5cm,angle=-90]{rmran.ps}
\includegraphics[width=5.5cm,angle=-90]{sf_mc.ps}
\caption{PDFs and structure functions of the patches HN and MN at high and 
medium latitudes. See Tables~\ref{patch} and \ref{statistic_par} for details of
 the patch parameters.}
\label{pdfsf_hnmn}
\end{figure}

It would take a huge amount of computing time to obtain the structure 
functions from all pixels of our simulated maps. We simplified this by randomly 
selecting $N_s$ pixels to conduct the calculation for structure functions. 
These pixels are randomly distributed across the maps. To accurately assess 
the structure functions for small angular separations, the number $N_s$ should 
be as large as possible. As a compromise we selected $N_s=2\times10^5$ pixels 
from the maps. The structure functions are shown in Figs.~\ref{pdfsf_hnmn}, 
\ref{pdfsf_ms}, \ref{pdfsf_pn} and \ref{pdfsf_hmp} (lower panels).  

\begin{table}
\setlength{\tabcolsep}{1.05mm}
\caption{Fitting results for structure functions and autocorrelation functions 
of RMs.}
\label{fit_par}
\centering
\begin{tabular}{l|rrrr|rrrr}\hline\hline
     & \multicolumn{4}{c}{structure functions} & 
       \multicolumn{4}{c}{autocorrelation function} \\
     \cline{2-9}
name & $A$ & $\Delta A$ & $m$ & $\Delta m$ 
     & $B$ & $\Delta B$ & $s$ & $\Delta s$ \\\hline

HN    &    34.8&   0.4& 0.916& 0.006&     8.6&   0.1&  0.957&  0.008\\
MN    &    62.7&   0.2& 0.862& 0.002&    15.7&   0.2&  0.853&  0.010\\
PNAC1 &  2951.1&  29.9& 0.719& 0.007&   596.0&   3.9&  0.773&  0.013\\
PSAC  &  2830.4&  11.6& 0.751& 0.002&   596.8&   5.9&  0.831&  0.017\\
MS    &    64.9&   0.5& 0.871& 0.005&    24.5&   0.4&  0.915&  0.010\\
HS    &    42.0&   0.4& 0.909& 0.005&    11.5&   0.1&  0.913&  0.006\\
PNC1  & 13695.7& 293.8& 0.658& 0.010&  2255.7&  16.4&  0.614&  0.018\\
PNC2  & 31230.9& 402.2& 0.676& 0.008&  5792.6&  40.0&  0.661&  0.013\\
PNAC2 &  2836.0&  57.4& 0.660& 0.013&   738.3&   5.5&  0.743&  0.014\\\hline
\end{tabular}
\end{table}

The theoretical deduction how structure functions are related to the 
power spectrum of magnetic fields has been given by \citet{ms96}, where the 
structure function of RMs is found to follow a power law as 
$D_{\rm RM}(\delta\theta)=A\delta\theta^m$ with $A$ being a constant and the 
spectral index $m$ being related to the power low index $\alpha$ of $P(k)$ as 
$m=-\alpha-2$. This relation holds for the inertial range, i.e.,
 for the scale $l_{\rm i}<l<l_{\rm o}$, where $l_{\rm i}$ and $l_{\rm o}$ are 
inner and outer scales, respectively. The angular inertial scale can thus be 
defined as $\delta\theta_{\rm i}<\delta\theta<\delta\theta_{\rm o}$, where 
$\delta\theta_{\rm i}=l_{\rm i}/r_{\rm max}$ and 
$\delta\theta_{\rm o}=l_{\rm o}/r_{\rm max}$. In principle the inertial range 
can be estimated by checking where the linear fit of the structure functions 
for a logarithmic scale fails. For the structure functions from our simulated 
results, we are able to infer the outer scale but the inner scale is smaller 
than the pixel size. Below we name the outer angular scale 
$\delta\theta_{\rm o}$ as the transition angle. The fit results for structure 
functions are listed in Table~\ref{fit_par}. 

\begin{figure}[!htbp]
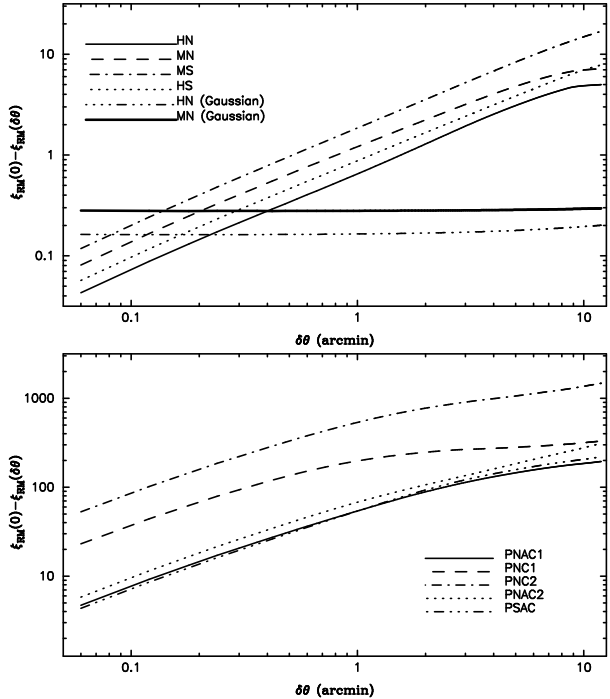

\centering
\includegraphics[angle=-90,width=8cm]{acf_l.ps}
\includegraphics[angle=-90,width=8cm]{acf_b.ps}
\caption{Autocorrelation functions for RMs.}
\label{acf}
\end{figure}

\citet{ev03} have noted that the very inner part of the autocorrelation 
function of RMs ($\xi_{\rm RM}$) can be written as 
$\xi_{\rm RM}(0)-\xi_{\rm RM}(\delta\theta)=B \delta\theta^s$, where $B$ is a 
constant, and $s=-\alpha-2$ is the spectral index of the magnetic energy 
spectrum. This implies that the power law index of the structure function  
$m$ and the power law index of the autocorrelation function $s$ should be 
equal. We obtain the autocorrelation functions by a Fourier transform of the 
RM power spectra and show them in Fig.~\ref{acf}. The RM fit results for each 
patch are listed in Table~\ref{fit_par}. As can be seen, the spectral 
indices of the autocorrelation functions do not deviate very much from that of 
the structure functions. We account the differences mainly to computational 
effects when obtaining the autocorrelation functions. Unwanted components might 
be introduced during  the pixelization of the maps when Fourier transformed. 
Therefore we are confident that the results based on the two methods are 
basically consistent with each other.

The spectral index of 5/3 is expected from the power law fit of the RM 
structure functions for a Kolmogorov-like turbulent magnetic field. This, 
however, is not seen in Table~\ref{fit_par}. Possible explanations are 
discussed in Sect.~4.7.1.

\subsection{Gaussian versus Kolmogorov-like random magnetic fields}

In the simulations presented by \citet{srwe08}, the random magnetic field 
components were realized {\it in situ} for each volume unit by Gaussian random 
numbers. The random fields created that way are called 'Gaussian fields' 
below. In that scheme, the simulated RMs are totally uncorrelated between the
 map pixels, which implies that no small-scale structures are expected. When 
random magnetic fields with a Kolmogorov-like spectrum are generated, as 
described in Sect.~\ref{b_realization}, a large number of turbulent boxes are 
placed along the line-of-sight. For arcsec angular resolutions several 
neighbouring line-of-sights pass the same boxes, which results in extended 
coherent features in the simulated RM maps.   

For the northern patches HN and MN, we also performed simulations with Gaussian 
 random magnetic field component for comparison. Both RM maps are shown in the 
first two rows in Fig.~\ref{rmmaps} and prove the above expectations. As seen 
from Fig.~\ref{pdfsf_hnmn}, the PDFs of the RM simulations with Gaussian random
 magnetic fields nicely conform to Gaussian distributions, while the PDFs from 
the simulations with Kolmogorov-like random magnetic fields do not. 
Quantitatively this can be verified by the skewness and kurtosis in 
Table~\ref{statistic_par}, where the values of these two parameters are much 
larger for Kolmogorov-like magnetic field spectra than Gaussian ones. The 
structure functions shown in the bottom panel of Fig.~\ref{pdfsf_hnmn} as well 
as the autocorrelation functions in Fig.~\ref{acf} reinforce the argument. The 
structure functions for Gaussian random magnetic fields are flat, while the 
structure functions for a Kolmogorov-like magnetic field spectrum have slopes, 
which will be discussed below. 

\subsection{RM patches from different realizations}

The RM structures in the present simulations are no predictions for sky 
emission structures towards the quoted directions, but are just one possible
 realization of the random magnetic fields following a Kolmogorov-like power 
law spectrum in context with other model parameters as described before. 

We generate the random magnetic field components in the following way. Their 
amplitudes are determined according to the slope of the power-law spectrum 
with a spectral index of $-11/3$, and the phase angles are realized by random 
numbers. The rotation angle of the numerous 10~pc boxes placed along the 
line-of-sight are also determined from random numbers. These random numbers are
 actually pseudo random and are initiated by an integer called ``seed". When 
the random ``seed" number is changed, a different realization of the same 
random magnetic field distribution is obtained.        

\begin{figure}[!htbp]
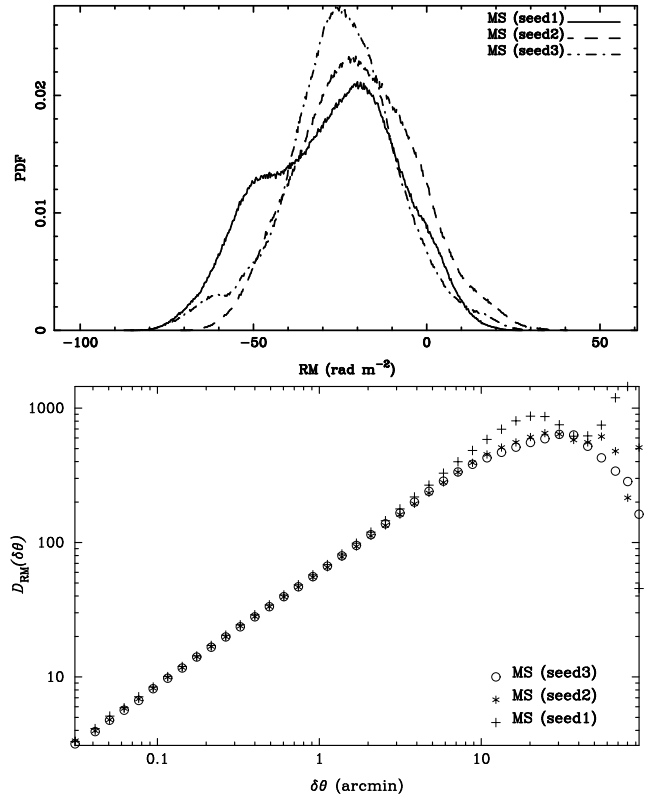

\centering
\includegraphics[width=5cm,angle=-90]{rmseed.ps}
\includegraphics[width=5.5cm,angle=-90]{sf_r.ps}
\caption{PDFs and structure functions of the patch MS for the same amplitudes 
but different realizations of the phases of the random magnetic field 
components created in a 10~pc box (see text for details).}
\label{pdfsf_ms}
\end{figure}

To check structure variations and structure functions for different 
realizations, we ran two more simulations for patch MS with different ``seed" 
numbers. The RM maps are displayed in the fifth row of Fig.~\ref{rmmaps} and 
show different distributions of RM features as expected. Also the PDFs for the 
RM maps (Fig.~\ref{pdfsf_ms}) vary considerably, as well as the parameters 
listed in Table~\ref{statistic_par}. The structure functions of RMs, however, 
are almost identical as shown in the bottom panel of Fig.~\ref{pdfsf_ms}, 
except for regions with large angular separations exceeding about $10\arcmin$, 
where the structure function becomes less accurate due to the decreasing number
of selected pixels.

The mean RM for the three realizations of patch MS is between 
$-$20~rad~m$^{-2}$ and $-$28~rad m$^{-2}$ (Table~\ref{statistic_par}) and the 
majority of RMs has negative values. However, due to the variance of up to 
19~rad~m$^{-2}$ (Table~\ref{statistic_par}) all three simulations show a number
 of small extended features in the maps with positive sign. Naively this could 
be interpreted as reversals of the large-scale magnetic field in case the 
available number of observed RMs from EGSs is small. However, in contrast with 
large-scale reversals in the Galactic plane as discussed for instance by 
\citet{hml+06}, the reversals seen here are entirely caused by the distribution
 of random magnetic fields. For this medium-latitude area, we roughly estimate 
the probability for picking up a ``reversal" to be about 7\%. This should be 
reflected in the distribution of observed RMs of EGSs and will prove the 
relevance of the simulations and the parameters used.  

\subsection{Galactic plane RM patches at different longitudes}

In this section we investigate RM variations in the plane versus Galactic 
longitudes. The four simulated patches have latitudes of about $+1\fdg2$. Two 
of them (PNC1 and PNC2) are in the inner Galaxy direction and the other two 
(PNAC1 and PNAC2) are in the anti-centre direction (see Table~\ref{patch}). 

\begin{figure}[!htbp]
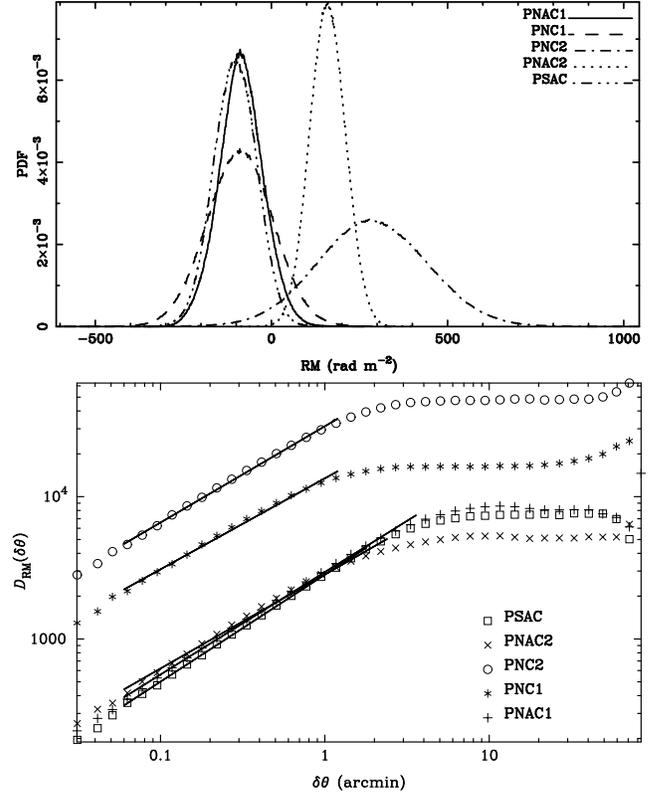

\centering
\includegraphics[width=5cm,angle=-90]{rml.ps}
\includegraphics[width=5.5cm,angle=-90]{sf_rm_b.ps}
\caption{PDFs and structure functions of the five simulated Galactic plane 
patches (see Table~\ref{patch} for field parameters).}
\label{pdfsf_pn}
\end{figure}

The visible structures in the maps (the third row in Fig.~\ref{rmmaps}) are 
less extended compared to those at higher latitudes. There is more fine 
structure for the inner Galaxy regions than for those in the anti-centre 
direction. The PDFs of all four simulated patches closely resemble Gaussian 
shapes (Fig.~\ref{pdfsf_pn}). The inner Galaxy patches PNC1 and PNC2 have 
smaller skewness values than the anti-centre patches 
(Table~\ref{statistic_par}), indicating that their PDFs are more close to 
Gaussian shapes. Note that the skewness and kurtosis for all patches is 
very small, which indicates that it is impossible to infer whether the 
turbulence spectrum is Gaussian or Kolmogorov-like based on their PDF. The 
structure functions as discussed before, however, clearly reveal the properties 
of the underlying turbulence spectrum. The average RM values are consistent 
with the RM results obtained by \citet{srwe08}.      

As displayed in Fig.~\ref{pdfsf_pn}, all Galactic plane structure functions are
 well fitted by power laws. Compared to the medium- and high-latitude 
fields a flattening of the structure functions occurs at smaller angular 
distances. The fit results for the power laws are listed in 
Table~\ref{fit_par}. From the parameters listed in Table~\ref{patch} we obtain 
a transition angle of about $1\arcmin$ for PNC1 and PNC2 and about $3\arcmin$ 
for PNAC1 and PNAC2. Fig.~\ref{pdfsf_pn} shows instead of a sharp change in the 
structure function a smooth transition from a linear slope to a flat curve 
in the logarithmic scale. The turn over of the structure functions ranges from 
about $1\arcmin$ to $2\arcmin$ for patches PNC1 and PNC2, from $3\arcmin$ to 
$5\arcmin$ for patch PNAC1, and from $2\arcmin$ to $5\arcmin$ for patch PNAC2, 
which are roughly consistent with the aforementioned expectations. The spectral
 indices are consistent with theoretical predictions by \citet{ms96} for 2D 
turbulence, which results in a spectral index of about 0.7 for the structure 
functions. However, our input turbulent magnetic fields are for 3D. The 
amplitudes of the structure functions increase towards the inner Galaxy 
regions, which is expected as the line-of-sights pass more turbulent cells or 
magnetic field boxes than towards the anti-centre or high-latitude direction. 
The amplitude of the structure function for PNC1 is larger than that for PNC2, 
which is mainly due to a larger electron density in this direction according 
to the NE2001 model.  

\subsection{RM patches at different latitudes}

\begin{figure}[!htbp]
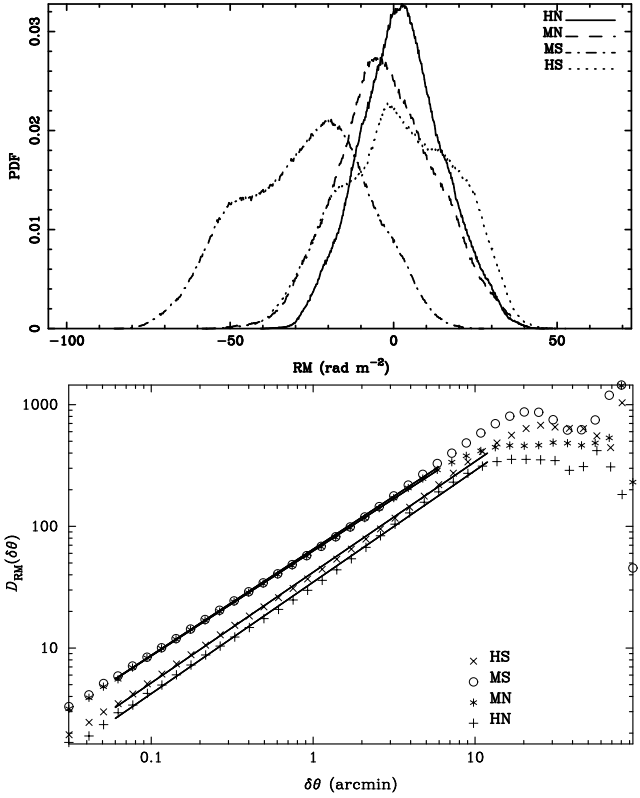

\centering
\includegraphics[width=5cm,angle=-90]{rmb.ps}
\includegraphics[width=5.5cm,angle=-90]{sf_rm_l.ps}
\caption{PDFs and structure functions of four patches (see first column of 
Fig.~\ref{rmmaps}) at roughly the same Galactic longitude located at medium and 
high latitudes (see Table~\ref{patch}).}
\label{pdfsf_hmp}
\end{figure}

We finally compare the different results towards different latitudes. The 
patches HN/HS, MN/MS,and PNAC1/PSAC have longitudes between $130\degr$ and 
$140\degr$. The latitudes are $\pm70\degr$, $\pm44\degr$ and $\pm 1\degr$. 
For patch MS the ``seed1" case is used. The six simulated RM maps are all shown
 in the first column of Fig.~\ref{rmmaps}. A clear trend for more extended 
structures for high-latitude regions is clearly seen. The PDFs for the 
high-latitude regions (HN/HS and MN/MS) all deviate significantly from Gaussian 
(Fig.~\ref{pdfsf_hmp} and Table~\ref{statistic_par}). 

The structure functions derived for the four high-latitude patches are shown in
Fig.~\ref{pdfsf_hmp} and for the five plane patches in Fig.~\ref{pdfsf_pn}. All
 of them have similar properties. The transition angle is calculated to be 
about $10\arcmin$ for HN/HS and MN/MS regions, which is consistent with the 
results displayed in Fig.~\ref{pdfsf_hmp}. The amplitudes of the structure 
functions are smaller for the high-latitude patches compared to those in the 
plane. This can be understood, because the length along the line-of-sight is 
small compared to that in the Galactic plane. The number of turbulent cells 
passed by the line-of-sights is smaller and hence is the amplitude of the 
structure functions. We also note that the slopes of the structure functions 
increase for the high-latitude patches. These spectral changes will be 
discussed below. 

\subsection{Discussion of RM simulations}

\subsubsection{Theoretical structure functions revisited}

From the structure functions of simulated RM maps, we found evidence for 
latitude-dependent variations of their spectral slope, which gets steeper 
towards high latitudes. The regular magnetic field and the electron density 
have spatial irregularities, which are of larger scale than the sizes of the 
patches. Therefore the large-scale magnetic field cannot cause variations of 
the spectral slopes, but influences the amplitude of the structure functions. 
The coupling between the electron density and random magnetic fields introduced 
by \citet{srwe08} augments the scattering of RMs and hence the amplitude of the 
structure function rather than changes its spectral index. For the present 
simulations, the random magnetic field is comparable to the regular magnetic 
field strength and dominates the regular field towards high-latitude regions. 
We conclude that all RM structures across the simulated patches are caused 
by the random magnetic field properties.  

\begin{table}[!htbp]
\caption{Fit parameters for structure functions from different simulation 
settings.}
\label{checkalpha}
\centering
\begin{tabular}{rrrrr}\hline\hline
\multicolumn{5}{c}{$L=20$~pc, $r_{\rm max}=10$~kpc}\\\hline
 $\Delta$ (pc) & $C$ & $\Delta C$ & $m$  & $\Delta m$ \\\hline 
    0.80  & 3.304 &  0.028 &  0.727 &   0.044\\
    0.40  & 3.430 &  0.021 &  0.644 &   0.033\\
    0.20  & 3.387 &  0.030 &  0.760 &   0.047\\
    0.10  & 3.414 &  0.009 &  0.612 &   0.014\\
    0.05  & 3.447 &  0.018 &  0.622 &   0.028\\\hline
\\
\multicolumn{5}{c}{$\Delta=0.5$~pc, $r_{\rm max}=10$~kpc} \\\hline
 $L$ (pc) & $C$ & $\Delta C$ & $m$  & $\Delta m$ \\\hline 
  200     & 2.830 & 0.006 &  1.262 &   0.006\\ 
  100     & 3.045 & 0.011 &  1.081 &   0.015\\
  50      & 3.201 & 0.008 &  0.982 &   0.010\\
  25      & 3.401 & 0.007 &  0.709 &   0.012\\\hline
\\
\multicolumn{5}{c}{$\Delta=0.25$~pc, $L=100$~pc} \\\hline
 $r_{\rm max}$ (kpc) & $C$ & $\Delta C$ & $m$  & $\Delta m$ \\\hline 
  0.01 &  $-$6.270 &  0.0181 & 1.660 &  0.014\\
  0.05 &  $-$4.243 &  0.0106 & 1.413 &  0.008\\
  0.10 &  $-$2.689 &  0.0128 & 1.668 &  0.009\\
  0.20 &  $-$1.742 &  0.0172 & 1.355 &  0.013\\
  0.40 &  $-$1.206 &  0.0202 & 1.689 &  0.013\\ 
  0.80 &     0.368 &  0.0133 & 1.282 &  0.012\\
  1.60 &     1.630 &  0.0088 & 1.330 &  0.009\\
  3.20 &     2.711 &  0.0060 & 1.154 &  0.006\\
  6.40 &     2.970 &  0.0083 & 1.151 &  0.010\\
 10.00 &     3.083 &  0.0078 & 1.081 &  0.010\\\hline 
\end{tabular}
\end{table}

The variations versus latitude must be ascribed to geometric effects, which 
reflects in the integral length ($r {\rm max}$) as listed in Table~\ref{patch}.
 The size of the magnetic field box ($L$) is fixed in the simulations 
presented, therefore, the integral length determines the number of turbulent 
cells or boxes passed by the line-of-sight. For a sufficient large number of 
cells, the fluctuations along the line-of-sight are averaged out. This results 
in a spectral slope of $m=-\alpha-3=2/3$ for the structure function. Note that 
this spectral index was observed and interpreted as a 2D turbulence in the 
interstellar medium by \citet{ms96}. 2D turbulence is confined to thin sheets 
or filaments. However, the present simulations are based on a 3D distribution 
of the turbulent magnetic field components, which have not the shape of sheets 
or filaments. When the number of cells is small enough so that the integral 
length is comparable to the outer scale, the structure function tends to have 
the index of $m=-\alpha-2=5/3$ as derived by \citep{ms96} for 3D 
Kolmogorov-like turbulence. Therefore we conclude that in the inertial range 
($\delta\theta_{\rm i}<\delta\theta<\delta\theta_{\rm o}$) the derived 
structure functions shall behave like 
$D_{\rm RM}(\delta\theta)\propto\delta\theta^m$ with $2/3<m<5/3$ depending on 
the number of cells. When the angular separation is larger than the transition 
angle, the structure function spectrum flattens and $m$ gradually turns to 
zero.

To prove these arguments we made several simulations for a patch of about 
$6\degr\times6\degr$ in the plane with $N_{\rm SIDE}=2048$ corresponding to 
a resolution of about $1\farcm7$. Although the resolution is much lower than 
the high-resolution simulations for the patches in Table~\ref{patch}, the 
dependence of the structure functions should be the same. These simulations can 
be realized in much smaller computing time than the arcsec resolution
simulations. We fit the structure functions by using 
$\log D_{\rm RM}(\delta\theta)=C+m\log\delta\theta$ and list the results 
including errors of C and m in Table~\ref{checkalpha}. We first fix the outer 
scale ($L$) and the integral length and vary the size of the small cubes 
($\Delta$). The results in Table~\ref{checkalpha} show that the size of the 
small cubes has little influence on the structure function parameter. We then 
change the outer scale, but keep the size of the small cubes and also the 
integral length constant. As expected the spectral index of the structure 
functions tends to be about 0.7 for very small outer scales, which means the 
number of turbulent cells is large. Finally, we increase the integral length 
from 10~pc to 10~kpc by using the same values of cube size and outer scale. 
The spectral index of the structure functions varies between 1.3 and 1.7 until 
the integral length exceeds about 1600~pc, becoming smaller for larger lengths. 
These test simulations support our conclusion that the spectral index of the 
structure function primarily depends on the number of turbulent cells, which is 
mainly determined by the length of the line-of-sight. Thus, when assessing the 
turbulent magnetic field properties from the structure functions of RMs, the 
line-of-sight dependence must be taken into account. In the Galactic plane the 
integral length is always large, which assures sufficient turbulent cells being 
passed through by the line-of-sight. Therefore the relation $\alpha=-m-3$ 
holds in general. Towards higher Galactic latitudes, however, the length of the 
line-of-sight decreases and hence also the number of turbulent cells decreases. 
The spectral index of the random fields approaches $\alpha=-m-2$. 

\subsubsection{Diagnosis of the Galactic foreground}

It has been proposed to conduct a very dense RM survey of EGS with the SKA to 
study magnetic field structures in the Milky Way, in galaxies, in cluster of 
galaxies and beyond \citep{bg04}. The key issue to achieve relevant information
 for extragalactic objects is to properly extract and separate the foreground 
RM contributions from the observations. We have tested the minimum number 
of pixel needed to calculate the correct structure function for the simulated 
maps and found that about 10000 pixel are required. For our map size this 
corresponds to a mean angular separation of about $1\arcmin$ in case the EGS 
are not larger than the pixel size of $1\farcs6$ and their intrinsic RMs 
average out.

In the plane the PDFs are close to Gaussian. Therefore the average of RMs is a 
good estimate of the foreground contribution caused by the large-scale magnetic 
field and the thermal electron distribution as modelled by \citet{srwe08}. 
However, large variances (Table~\ref{statistic_par}) and amplitudes of the 
structure functions (Table~\ref{fit_par}) imply that after having subtracted 
the average RM in a certain direction the RM scattering of the foreground 
emission is still preserved. Thus patches in the plane are certainly not well 
suited for the study of extragalactic magnetic fields.
   
At high latitudes the PDFs deviate significantly from a Gaussian distribution
and also more extended structures show up (Fig.~\ref{rmmaps}). To take the 
foreground RMs appropriately into account a dense sample of EGS is required. 
In case an extended extragalactic object is seen towards an excessive RM patch 
of the same size, additional information is required to solve the RM foreground
 problem. On the other hand, the variation of the RMs are much smaller than in 
the plane. Thus high-latitude regions seem better suited to study extragalactic 
magnetic fields. However, according to \citet{kbm+08} cosmological magnetic 
fields of a few $10^{-7}$~$\mu$G at a redshift of 2 have an expected RM excess 
of up to 20~rad~m$^{-2}$. This is of the same order of magnitude as the RM 
variance of the Galaxy at high latitudes obtained from the present simulations, 
which means a detection of cosmological magnetic fields is quite challenging. 
Note that reducing the coupling factor between electron density and random 
fields introduced by \citet{srwe08} to explain the observed depolarization of 
Galactic synchrotron emission at low angular resolution will in turn also 
reduce the foreground RM variance. Therefore future high-resolution 
observations of the small-scale structures of the magnetized ISM are needed to 
investigate the coupling factor in detail. Such observations are planned to be 
carried out with LOFAR, which will operate at m-wavelengths and will thus be 
able to detect very small RM-variations from small-scale Galactic emission 
features.  

\begin{table*}[!htbp]
\setlength{\tabcolsep}{1.4mm}
\caption{Statistical parameters for the simulated $I$, $PI$, and $PA$ patches. 
The patch designations are listed in Table~\ref{patch}.}
\label{statistic_pari}
\centering
\begin{tabular}{l|rrrr|rrrr|rrrr}\hline\hline
           & \multicolumn{4}{c}{$I$}  & \multicolumn{4}{c}{$PI$}  & \multicolumn{4}{c}{$PA$}  \\
           \cline{2-13}
           & average  & variance &  skewness & kurtosis &  average  & variance &  skewness & kurtosis &  average  & variance  &  skewness & kurtosis \\ 
name       &  (mK)    &  (mK)    &           &          &   (mK)    &  (mK)    &           &          & ($\degr$) & ($\degr$) &           &          \\\hline
HN         &  451.53  &  17.81   &   0.31  &$-$0.049 &  48.59    & 13.63    &$-$0.04  &$-$0.347 &   32.85   &    8.81   & $-$0.36 &   0.191 \\
HN (Gaussian)&  446.73  &  13.98   &   0.06  &   0.003 &  57.91    & 10.49    &   0.03  &   0.003 &   25.04   &    6.15   & $-$0.02 &   0.122 \\
MN         &  489.76  &  14.19   &$-$0.39  &$-$0.007 &  48.60    & 13.33    &$-$0.13  &   0.052 &   29.31   &    9.45   &    0.07 &   1.599 \\
MN (Gaussian)&  495.20  &  14.67   &   0.06  &   0.004 &  63.96    & 11.86    &$-$0.09  &   0.094 &   16.09   &    8.07   & $-$0.05 &   0.174 \\
PNAC1      & 1557.63  &  36.62   &$-$0.26  &$-$0.052 & 106.32    & 39.35    &   0.13  &$-$0.226 & $-$9.15   &   14.80   &    0.03 &   2.527 \\
PSAC       & 1630.36  &  39.12   &$-$0.28  &   0.233 & 111.65    & 50.87    &   0.19  &$-$0.554 &$-$18.93   &   23.68   &    0.54 &   2.327 \\
MS (seed1) &  519.56  &  18.59   &   0.46  &$-$0.079 &  85.53    & 25.60    &$-$0.78  &$-$0.101 &$-$35.46   &   18.62   &    0.82 &   4.854 \\
MS (seed2) &  540.35  &  16.87   &   0.18  &   0.873 & 122.55    & 25.25    &$-$0.65  &   0.170 &$-$29.04   &   14.70   & $-$0.02 &$-$0.686 \\
MS (seed3) &  556.77  &  19.11   &$-$0.09  &$-$0.563 &  87.20    & 19.04    &$-$0.12  &$-$0.296 &$-$37.31   &   11.56   &    0.39 &$-$0.202 \\
HS         &  512.79  &  24.02   &   0.25  &$-$0.492 &  78.68    & 22.62    &$-$0.15  &$-$0.290 &$-$24.73   &   21.44   &    0.04 &   0.336 \\
PNC1       & 2996.47  & 121.62   &   0.10  &$-$0.710 &  61.85    & 32.01    &   0.59  &   0.140 &$-$14.66   &   50.31   &    0.54 &$-$0.850 \\
PNC2       & 2780.71  &  93.90   &   0.13  &$-$0.040 &  54.90    & 27.98    &   0.56  &   0.075 &   15.26   &   37.16   & $-$0.66 &   0.348 \\
PNAC2      & 1449.76  &  40.76   &   0.31  &   0.419 &  71.01    & 29.17    &   0.19  &$-$0.235 &   45.83   &   31.16   & $-$2.51 &   7.768 \\\hline 
\end{tabular}
\end{table*}

\begin{figure*}[!htbp]
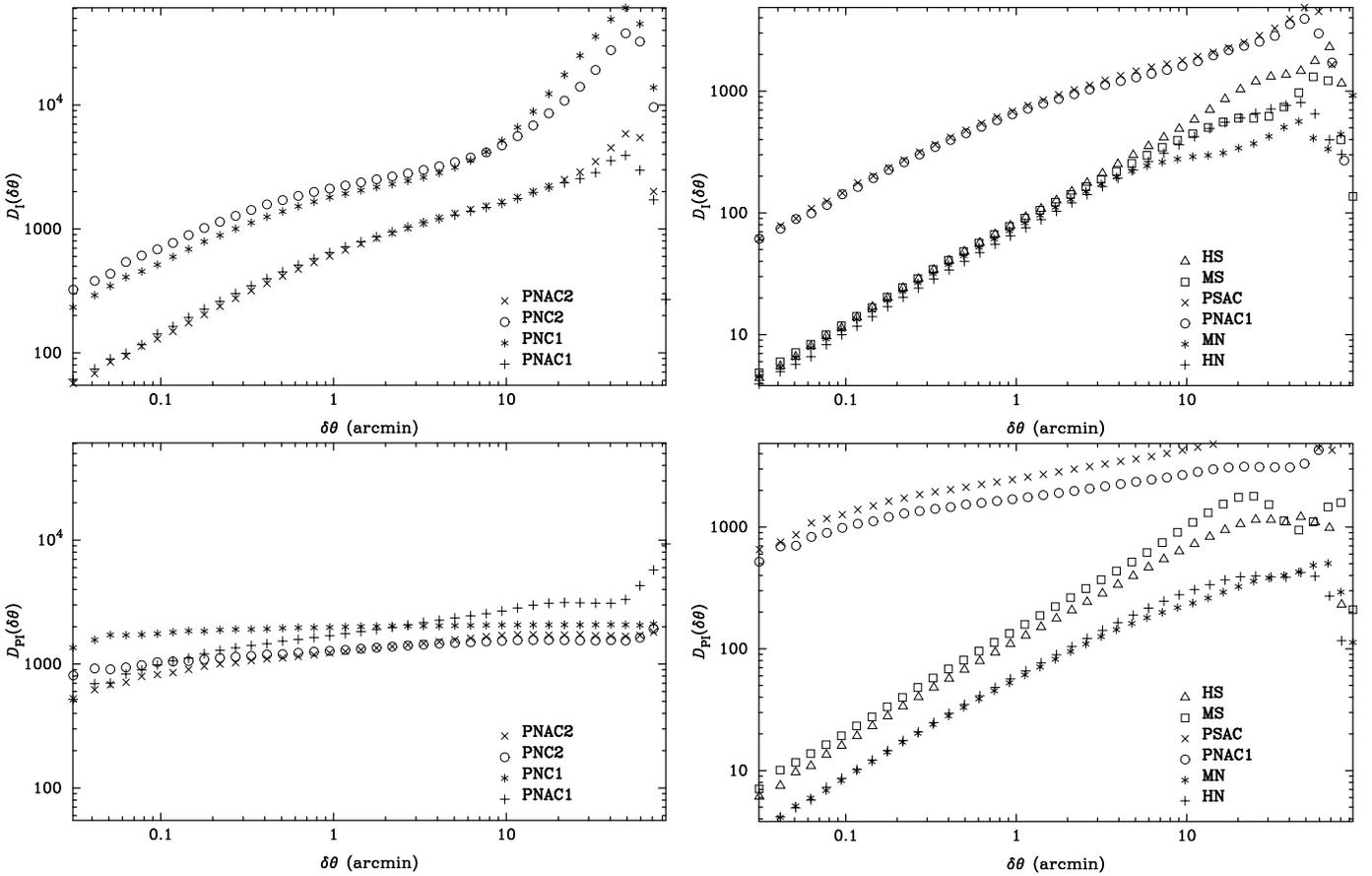

\begin{minipage}{0.5\textwidth}
\centering
\includegraphics[angle=-90,width=9cm]{sf_i_b.ps}
\end{minipage}
\begin{minipage}{0.5\textwidth}
\centering
\includegraphics[angle=-90,width=9cm]{sf_i_l.ps}
\end{minipage}
\begin{minipage}{0.5\textwidth}
\centering
\includegraphics[angle=-90,width=9cm]{sf_pi_b.ps}
\end{minipage}
\begin{minipage}{0.5\textwidth}
\centering
\includegraphics[angle=-90,width=9cm]{sf_pi_l.ps}
\end{minipage}
\caption{Structure functions for the total intensity $I$ (top panels) and the 
polarized intensity $PI$ maps (bottom panels).}
\label{sf_ipi}
\end{figure*}

\section{Simulated total intensity and polarization maps}

We briefly describe the properties of the simulated total intensity and linear 
polarization maps emerging from the same simulation runs where the already 
discussed RM distributions were calculated. These simulated Stokes $I$, $U$, 
$Q$ maps are all for the frequency of 1.4~GHz, whereas the RM distribution is 
frequency-independent. 

The simulated total intensity $I$, $U$, $Q$, polarized intensity $PI$ and 
polarization angle $PA$ maps are shown in Figs.~\ref{imaps}, \ref{umaps}, 
\ref{qmaps}, \ref{pimaps}, and \ref{pamaps} for the same patches and simulation 
variations as displayed in Fig.~\ref{rmmaps} for RMs. The statistical 
parameters for the $I$, $PI$, and $PA$ maps are calculated and listed in 
Table~\ref{statistic_pari}. The structure functions for $I$ and $PI$ are shown 
in Fig.~\ref{sf_ipi}.  

Similar as for RM maps, the patches in the plane are dominated by structures on 
very small scales, while the patches at high latitudes exhibit more extended 
features. These structures are all related to the properties of the turbulent 
magnetic fields. The variation of the sizes of the features is primarily due to 
the much larger integral length along the line-of-sights in the plane than at 
high latitudes.

The large-scale cosmic-ray and magnetic field also have imprints on the 
simulations. For example, the clear gradient of the total intensity 
particularly in the plane maps towards the inner Galaxy traces the large-scale 
magnetic field and the cosmic-ray distribution. The asymmetric halo field adds 
up to the disk field and enhances the field strength below the Galactic plane, 
which produces larger intensities there. This can be seen from the averages of 
$I$ and $PI$ intensities listed in Table~\ref{statistic_pari} for the PSAC and 
the PNAC1 patch. 

The average polarization percentage is about 10\% for the HN and MN patches and 
about 16\% for the HS and MS patches. Towards these directions both RM and its 
variance are small so that Faraday depolarization is small as well 
\citep{sbs+98}. The low degree of polarization is thus due to the random 
magnetic fields, which dominate the field components. The percentage becomes 
about 7\% for patch PNAC1, 5\% for patch PNAC2, and 2\% for the inner Galaxy 
patches, which is caused by strong Faraday depolarization, as both the RM and 
the RM scattering are large. There is no indication for a correlation among the 
variances of RM, polarized intensity $PI$ and polarization angle $PA$.       

The structure functions for total intensity $I$ (Fig.~\ref{sf_ipi}) have a 
linear slope in the logarithmic scale, but these are generally more 
shallow than those for the corresponding RMs. The total intensity $I$ and 
polarized intensity $PI$ maps at 
high latitudes have quite similar structure functions, while they are rather 
distinct from each other in the plane. The structure functions of $PI$ are 
nearly flat. This indicates that due to strong depolarization in the plane the 
$PI$ structure functions do not contain the fluctuation information of the 
turbulent magnetic field any more.    

\section{Conclusions}

Galactic 3D-models of the distribution of thermal electrons, cosmic-ray 
electrons, and magnetic fields were used to calculate total intensity, 
polarization and RM maps for selected patches of the Galaxy via the 
\textsc{hammurabi} code. The angular resolution of the maps is about 
$1\farcs6$, close to that of the planned SKA. The simulated patches are 
distributed at different longitudes and latitudes, which enables us to study 
spatial variations of the results. The simulated RM maps are of particular 
interest, as they are directly related to the planned deep RM EGS survey with 
the SKA. Complete SKA simulations need in addition simulated sky patches
with distributed polarized sources. Such simulations were made by 
\citet{wmj+08}.

In the present simulations the random magnetic field component is assumed to 
follow a Kolmogorov-like power spectrum. Compared to purely random Gaussian 
magnetic fields, structures on different scales show up in the RM as well as in 
the total intensity $I$ and the polarized intensity $PI$ maps. The simulations 
predict more extended structures for high-latitude regions. 

We study the PDFs and structure functions for the RM maps. The PDFs of the 
patches in the plane are close to Gaussian, whereas at high latitudes clear 
deviations from Gaussian shapes are noted. The structure functions have a 
power law slope for angular separations smaller than the transition angle 
and 
flatten for larger angular separations. The transition angle is determined by 
the outer scale of the turbulence and by the length of the line-of-sight. The 
amplitudes of the structure functions become smaller and the slopes of the 
structure functions steepen towards higher-latitude regions. All this is 
entirely caused by geometric effects. In the plane, the integral length is much 
larger than at high latitudes, therefore more turbulent cells are passed by the 
line-of-sights. This significantly smears out fluctuations along the 
line-of-sights, which makes the 3D turbulence observationally a 2D turbulence. 
Thus the structure functions are flatter in the plane than at high latitudes. 
We notice that the PDFs of RMs have a wide distribution in all directions, 
which potentially leads to depolarization of Galactic radio emission, when 
observed with larger beams. It also strongly influences the interpretation of 
discrete features sampled by RMs of EGS. It is obvious that only a very dense 
grid of extragalactic RMs will help to separate Galactic foreground influence 
from extragalactic contributions. This is planned for the SKA.  

The simulations at 1.4~GHz presented in this paper can be easily run at any 
other frequency. While the RM maps are frequency independent, total intensity 
and polarization simulations are of particular interest for low-frequency
simulations, where LOFAR is expected to do sensitive high-resolution
observations in the near future. At low frequencies, thermal absorption adds to 
depolarization effects and must be properly taken into account. LOFAR will 
make a major attempt to trace the very weak signal from the Epoch of 
reionization, where the selection of a high-latitude Galactic area with very 
low foreground signal is important for the success of these ambitious
low-frequency observations as well as the subtraction of all kinds of 
extragalactic foregrounds as discussed by \citet{jelic08}. 

\begin{acknowledgements}
X.~H.~Sun acknowledges support by the European Community Framework Programme 6,
Square Kilometre Design Study (SKADS), contract no 011938. We are very grateful
to Andr{\`e} Waelkens and Torsten En{\ss}lin for providing the 
\textsc{hammurabi} code and for many detailed discussions on its application, 
improvements and results.  We thank Michael Kramer and Patricia Reich for 
critical reading of the manuscript and helpful comments. We like to thank 
Walter Alef for support in using the MPIfR PC-cluster operated by the 
VLBI-group, where all \textsc{hammurabi} based simulations were performed.
\end{acknowledgements}

\bibliographystyle{aa}

\begin{thebibliography}{}
\bibitem[Armstrong et al.(1995)]{ars95}
Armstrong, J. W., Rickett, B. J., \& Spangler, S. R. 1995, ApJ, 443, 209 
\bibitem[Beck \& Gaensler(2004)]{bg04}
Beck, R., \& Gaensler, B. M. 2004, New Astronomy Review, 48, 1289
\bibitem[Berkhuijsen et al.(2006)]{bmm06}
Berkhuijsen, E. M., Mitra, D., \& M\"uller, P. 2006, AN, 327, 82
\bibitem[Brown et al.(2003)]{btj03}
Brown, J. C., Taylor, A. R., \& Jackel, B. J. 2003, ApJS, 145, 213
\bibitem[Brown et al.(2007)]{bhg07}
Brown, J. C., Haverkorn, M., Gaensler, B. M., et al. 2007, ApJ, 663, 258
\bibitem[Cordes \& Lazio(2002)]{cl02}
Cordes, J. M., \& Lazio, T. J. W. 2002, preprint (arXiv: astro-ph/0207156)
\bibitem[En{\ss}lin \& Vogt(2003)]{ev03}
En{\ss}lin, T. A., \& Vogt, C. 2003, A\&A, 401, 835
\bibitem[Gaensler et al.(2008)]{gmcm08}
Gaensler, B. M., Madsen, G. J., Chatterjee, S., \& Mao, S. A. 2008, PASA, 25, 184
\bibitem[G\'orski et al.(2005)]{ghb+05}
G\'orski, K. M., Hivon, E., Banday, A. J., et al. 2005, ApJ, 622, 759
\bibitem[Han et al.(1999)]{hmq99}
Han, J. L., Manchester, R. N., \& Qiao, G. J. 1999, MNRAS, 306, 371
\bibitem[Han et al.(2006)]{hml+06}
Han, J. L., Manchester, R. N., Lyne, A. G., Qiao, G. J., \& van Straten, W. 
2006, ApJ, 642, 868
\bibitem[Haslam et al.(1982)]{hss+82}
Haslam, C. G. T., Salter, C. J., Stoffel, H., \& Wilson, W. E. 1982, A\&AS, 47, 1
\bibitem[Haverkorn et al.(2008)]{hbgm08}
Haverkorn, M., Brown, J. C., Gaensler, B. M., \& McClure-Griffiths, N. M. 2008, ApJ, 680, 362 
\bibitem[Hinshaw et al.(2007)]{hin07}
Hinshaw, G., Nolta, M. R., Bennett C. L., et al. 2007, ApJS, 170, 288
\bibitem[Jeli\'c et al.(2008)]{jelic08}
Jeli{\'c}, V., Zaroubi, S., Labropoulos, P., et al. 2008, MNRAS, 389, 1319   
\bibitem[Kowal et al.(2007)]{klb07}
Kowal, G., Lazarian, A., \& Beresnyak, A. 2007, ApJ, 658, 423
\bibitem[Kronberg et al.(2008)]{kbm+08}
Kronberg, P. P., Bernet, M. L., Miniati, F., Lilly, S. J., Short, M. B., \& 
Higdon, D. M. 2008, ApJ, 676, 70
\bibitem[Kronberg \& Perry(1982)]{kp82}
Kronberg, P. P., \& Perry, J. J. 1982, ApJ, 263, 518
\bibitem[Minter \& Spangler(1996)]{ms96}
Minter, A. H., \& Spangler, S. R. 1996, ApJ, 458, 194
\bibitem[Mitra et al.(2003)]{mwkj03}
Mitra, D., Wielebinski, R., Kramer, M., \& Jessner, A. 2003, A\&A, 398, 993
\bibitem[Mizeva et al.(2007)]{mrf+07}
Mizeva, I., Reich, W., Frick, P., Beck, R., \& Sokoloff, D. 2007, AN, 328, 80
\bibitem[Moss \& Sokoloff(2008)]{ms08}
Moss, D., \& Sokoloff, D. 2008, A\&A, 487, 197
\bibitem[Noutsos et al.(2008)]{njkk08}
Noutsos, A., Johnston, S., Kramer, M., \& Karastergiou, A. 2008, MNRAS, 386, 1881
\bibitem[Page et al.(2007)]{phk+07}
Page, L., Hinshaw, G., Komatsu, E., et al. 2007, ApJS, 170, 335
\bibitem[Simonetti et al.(1984)]{scs84}
Simonetti, J. H., Cordes, J. M., \& Spangler, S. R. 1984, ApJ, 284, 126
\bibitem[Sokoloff et al.(1998)]{sbs+98}
Sokoloff, D. D., Bykov, A. A., Shukurov, A., Berkhuijsen, E. M., Beck, R., \&
Poezd, A. D. 1998, MNRAS, 299, 189
\bibitem[Sun \& Han(2004)]{sh04}
Sun, X. H., \& Han, J. L. 2004, in "The Magnetized Interstellar Medium",
eds. B. Uyan{\i}ker, W. Reich \& R. Wielebinski, Copernicus GmbH, p.25
\bibitem[Sun et al.(2008)]{srwe08}
Sun, X. H., Reich, W., Waelkens, A., \& En{\ss}lin, T. A. 2008, A\&A, 477, 573 
\bibitem[Taylor et al.(2007)]{tsj+07}
Taylor, A. R., Stil, J. M., Grant, J. K., et al. 2007, ApJ, 666, 201 
\bibitem[Testori et al.(2008)]{trr08}
Testori, J. C., Reich, P., \& Reich, W. 2008, A\&A, 484, 733 
\bibitem[Waelkens(2005)]{wae05}
Waelkens, A. 2005, Diploma Thesis, Ludwig-Maximilian-Universit\"at M\"unchen
\bibitem[Waelkens et al.(2009)]{wjr+09}
Waelkens, A., Jaffe, T., Reinecke, M., Kitaura, F. S., \& En{\ss}lin, T. A. 2009, A\&A, 495, 697 
\bibitem[Wilman et al.(2008)]{wmj+08}
Wilman, R. J., Miller, L., Jarvis, M. J., et al. 2008, MNRAS, 388, 1335
\bibitem[Wolleben et al.(2006)]{wlrw06}
Wolleben, M., Landecker, T. L., Reich, W., \& Wielebinski, R. 2006, A\&A, 448, 441 
 
\end{thebibliography}

\appendix

\section{Details for the realization of random magnetic field components}
\label{b_fft}
The general concept is to generate random magnetic fields in a box by a Fast 
Fourier Transformation (FFT), randomly rotate the box around its centre  to 
obtain a different view on the box, and finally fill the volume of the Galaxy with these 
randomly rotated boxes. The size of a box is $L^3$ and the size of the Galaxy is 
$D_X\times D_Y\times D_Z$. 

We first define several Cartesian coordinate systems:
\begin{itemize}
\item $(X,\,Y,\,Z)$: with $XY$-plane parallel to the Galactic plane and the
                     Galactic centre located at $(D_X/2,\,D_Y/2,\,D_Z/2)$. 
\item $(x,\,y,\,z)$: based on the box with the origin at its vertex. 
                     When the Galaxy is filled with boxes the origin of the 
                     first box is the same as for the system $(X,\,Y,\,Z)$ and
                     the $xy$-plane is parallel to the $XY$-plane. 
\item $(x^\prime,\,y^\prime,\,z^\prime)$ or  
      $(x^{\prime\prime},\,y^{\prime\prime},\,z^{\prime\prime})$: the same 
      as $(x,\,y,\,z)$, but with the origin at the centre of the box.
\end{itemize}

\begin{figure}[!htbp]
\includegraphics[width=9cm]{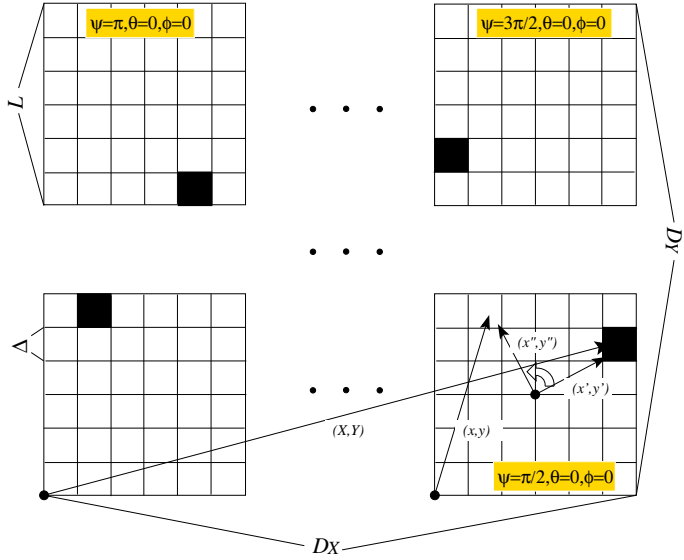}
\caption{2D-scheme for the realization of random magnetic fields. The sizes of 
the box and the cube are $L$ and $\Delta$, respectively. These boxes cover a 
region of $D_X\times D_Y$. The Euler angles, by which the boxes are rotated 
from the lower left box, are indicated. The shaded box marks the same cube 
after rotation. 
}
\label{bb}
\end{figure}

Based on these coordinate systems the whole process is illustrated in 
Fig.~\ref{bb}. 
 
We first consider the random magnetic field distribution in a box. The random 
magnetic field is assumed to be isotropic and homogeneous, and its three 
components do not correlate with each other. The three components thus follow 
the same power law spectrum $P(k)=C_b^2k^\alpha$. Here $C_b^2$ is the 
amplitude, $k=2\pi/l$ is the wave vector with $l$ being the spatial scale and 
$\alpha$ is the spectral index. For the simulations presented in this paper 
a Kolmogorov-like turbulence is assumed, which means $\alpha=-11/3$.

We first generate the 1D component of the random magnetic field. The other two 
components were obtained by simply repeating the process for the first one. 
This 1D component denoted as $b(x,\,y,\,z)$ is the inverse Fourier 
transformation of the power spectrum as the following,
\begin{equation}\label{ft}
b(\mathbf{r})=\!\!\int\!\!\!\sqrt{(P(k))}
\exp(i\xi(\mathbf{r})+i\mathbf{k}\cdot\mathbf{r})\,\,{\rm d}
\mathbf{r},
\end{equation}
where $\xi$ is the phase angle taken as a uniform random number between 0 and
$2\pi$, $i=\sqrt{-1}$ and $\mathbf{r}=(x,y,z)$ is the position vector. 
In practice Equation~(\ref{ft}) needs to be discretised. Therefore we divide 
the box into $N^3$ cubes with a size of $\Delta^3$ with $\Delta=L/N$ 
(Fig.~\ref{bb}). Equation (\ref{ft}) is then expressed as,   
\begin{equation}\label{bx}
b_{l^\prime m^\prime n^\prime}=\sum_{l=0}^{N-1}\sum_{m=0}^{N-1}
\sum_{n=0}^{N-1} \mathcal{P}_{lmn}\times 
\exp(i\Delta(k_ll^\prime+k_mm^\prime+k_nn^\prime)),
\end{equation}
where $\mathcal{P}_{lmn}=\sqrt{P_{lmn}}\exp(i\xi_{lmn})$,
$P_{lmn}=C_b^2k^{\alpha}_{lmn}$, and $k_{lmn}=\sqrt{k_l^2+k_m^2+k_n^2}$. 
Here
\begin{equation}
k_l=\left\{
\begin{array}{ll}
2\pi l/L     &\quad\quad {\rm for}\,\, l\leq N/2\\[3mm] 
2\pi(N-l)/L  &\quad\quad {\rm for}\,\, l>N/2,
\end{array}
\right.
\end{equation} 
with the same for $k_m$ and $k_n$. Generally $k_{lmn}$ is limited to 
the inertial range $2\pi/l_{\rm o}\leq k_{lmn}\leq2\pi/l_{\rm i}$, where 
$l_{\rm i}=\Delta$ and $l_{\rm o}=L$ are the inner scale and outer scale of 
the turbulence, respectively. To assure that $b_{l^\prime m^\prime n^\prime}$ 
is a real number $\mathcal{P}_{lmn}$ should be Hermitian, i.e. conforming to 
the following, 
\begin{itemize}
\item
   \begin{equation}
      \mathcal{P}_{lmn}=\overline{\mathcal{P}_{N-l\,N-m\,N-n}},
   \end{equation}
\item $\mathcal{P}_{lmn}$ is a real number, if the following is satisfied,   
   \begin{equation}
      \left\{
        \begin{array}{ll}
           l=m=n=0 & \quad {\rm mod}(N,2)=1 \\[3mm] 
           l=0,\,N/2,\, m=0,\,N/2,\, n=0,\,N/2  & \quad {\rm mod}(N,2)=0. 
        \end{array}
      \right.
   \end{equation}
\end{itemize}

The FFT package provided by the FFTW library \footnote{www.fftw.org} is
applied to obtain the random magnetic field distribution. 

To calculate the random magnetic field in any position ($X$, $Y$, $Z$) in the 
Galaxy $b(X,\,Y,\,Z)$, the following procedure is executed (Fig.~\ref{bb}),
\begin{enumerate}
 \item  transform $(X,\,Y,\,Z)$ to $(x^\prime,\,y^\prime,\,z^\prime)$ as, 
        \begin{equation}
           x^\prime=\frac{X}{L}-i_x L - \frac{L}{2},\,
           y^\prime=\frac{Y}{L}-i_y L - \frac{L}{2},\, 
           z^\prime=\frac{Z}{L}-i_z L - \frac{L}{2}
        \end{equation}
        where $i_x={\rm int}(X/L)$, $i_y={\rm int}(Y/L)$ and 
        $i_z={\rm int}(Z/L)$ are the integer part. 

 \item  rotate $(x^\prime,\,y^\prime,\,z^\prime)$ by Euler angles 
        $(\psi,\,\theta,\,\phi)$ around the centre of the box to 
        get $(x^{\prime\prime},\,y^{\prime\prime},\,z^{\prime\prime})$ as,
        \begin{equation}
           \left[
            \begin{array}{c}
              x^{\prime\prime}\\[1.5mm]
              y^{\prime\prime}\\[1.5mm]
              z^{\prime\prime}
            \end{array}
           \right]
           =\mathbf{B}\mathbf{C}\mathbf{D}
           \left[
             \begin{array}{c}
              x^\prime\\[1.5mm]
              y^\prime\\[1.5mm]
              z^\prime
             \end{array}
            \right]
        \end{equation}
        and
        \begin{equation}
          \mathbf{B}=\left[
             \begin{array}{ccc}
                 \cos\psi  &  \sin\psi   &  0 \\[1.5mm]
                 -\sin\psi &  \cos\psi   &  0 \\[1.5mm]
                 0         &  0          &  1
             \end{array}
           \right]
        \end{equation}
        \begin{equation}
          \mathbf{C}=\left[
            \begin{array}{ccc}
               1 &  0            &  0          \\[1.5mm]
               0 &  \cos\theta   &  \sin\theta \\[1.5mm]
               0 & -\sin\theta   &  \cos\theta
            \end{array}
                \right]
        \end{equation}
        \begin{equation}
           \mathbf{D}=\left[
             \begin{array}{ccc}
               \cos\phi  &  \sin\phi  &  0  \\[1.5mm]
               -\sin\phi &  \cos\phi  &  0  \\[1.5mm]
               0         &  0         &  1
             \end{array}
               \right]
        \end{equation}
        Here the Euler angles are obtained as,
        \begin{equation}
          \phi=\left\{
             \begin{array}{lcr@{\leq\mathcal{R}_1\!\!<}l}
               0              &\,\,   & 0     &  0.25 \\[1.5mm]
               \frac{\pi}{2}  &\,\,   & 0.25  &  0.5  \\[1.5mm]
               \pi            &\,\,   & 0.5   &  0.75 \\[1.5mm]
               \frac{3\pi}{2} &\,\,   & 0.75  &  1
             \end{array}
               \right.,
        \end{equation}
        \begin{equation}
           \theta=\left\{
            \begin{array}{lcr@{\leq\mathcal{R}_2\!\!<}l}
               0              &\,\,   & 0     &  0.33 \\[1.5mm]
               \frac{\pi}{2}  &\,\,   & 0.33   & 0.67 \\[1.5mm]
               \pi            &\,\,   & 0.67   & 1
            \end{array}
             \right.,
            \end{equation}
        and
        \begin{equation}
           \psi=\left\{
              \begin{array}{lcr@{\leq\mathcal{R}_3\!\!<}l}
                 0              &\,\,   & 0     &  0.25 \\[1.5mm]
                 \frac{\pi}{2}  &\,\,   & 0.25  &  0.5  \\[1.5mm]
                 \pi            &\,\,   & 0.5   &  0.75 \\[1.5mm]
                 \frac{3\pi}{2} &\,\,   & 0.75  &  1
              \end{array}
                \right..
        \end{equation}

        Here we use three random numbers $\mathcal{R}_1$, $\mathcal{R}_2$, and 
        $\mathcal{R}_3$ to determine the Euler angles. For simplicity the Euler
        angles are set to be integer multiple of $\pi/2$ to avoid complex 
        projections. 
  
 \item  transform $(x^{\prime\prime},\,y^{\prime\prime},\,z^{\prime\prime})$ 
        to $(x,\,y,\,z)$ then to $(l^\prime,\,m^\prime,\,n^\prime)$, 
        \begin{equation}
          x=x^{\prime\prime}+\frac{L}{2}, \quad
          y=y^{\prime\prime}+\frac{L}{2}, \quad
          z=z^{\prime\prime}+\frac{L}{2},
        \end{equation}
        and 
        \begin{equation}
          l^\prime={\rm int}(\frac{x}{\Delta}), \quad
          m^\prime={\rm int}(\frac{y}{\Delta}), \quad
          n^\prime={\rm int}(\frac{z}{\Delta}).
        \end{equation}
        Finally $b(X,\,Y,\,Z)=b_{l^\prime m^\prime n^\prime}$.
\end{enumerate}

\section{Simulated total intensity and polarization maps}
\label{othermaps}

\begin{figure*}[!htbp]
\centering
\includegraphics[width=18cm]{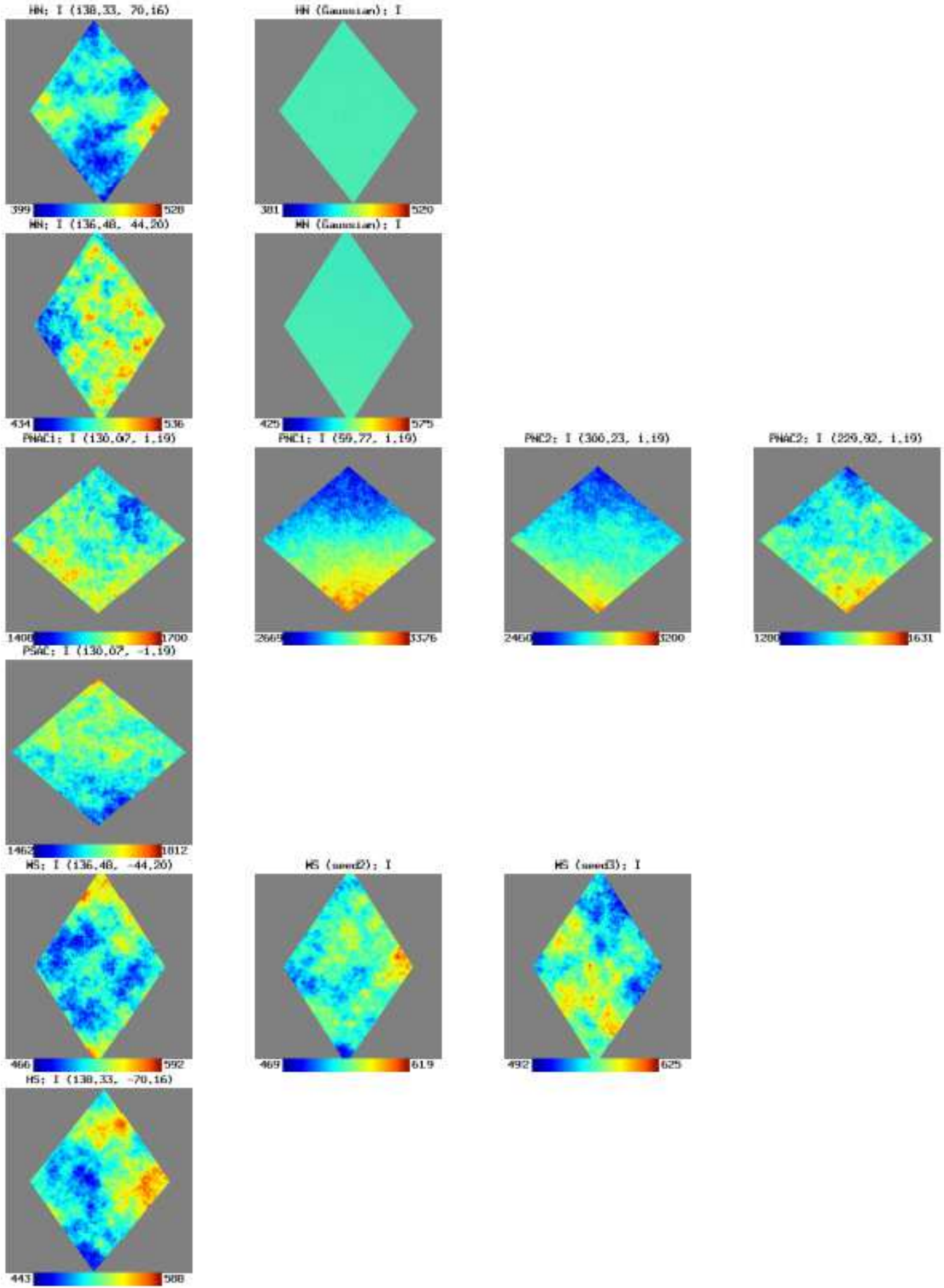}
\caption{Total intensity $I$ maps for all the simulated patches in gnomonic 
projection. The centre coordinates are given on top of each panel together with 
its label (see Table~\ref{patch} for details). Intensities are in mK brightness 
temperature at 1.4~GHz. The minimum and maximum of each map is shown together 
with the wedge below each panel.}
\label{imaps}
\end{figure*}

\begin{figure*}[!htbp]
\centering
\includegraphics[width=18cm]{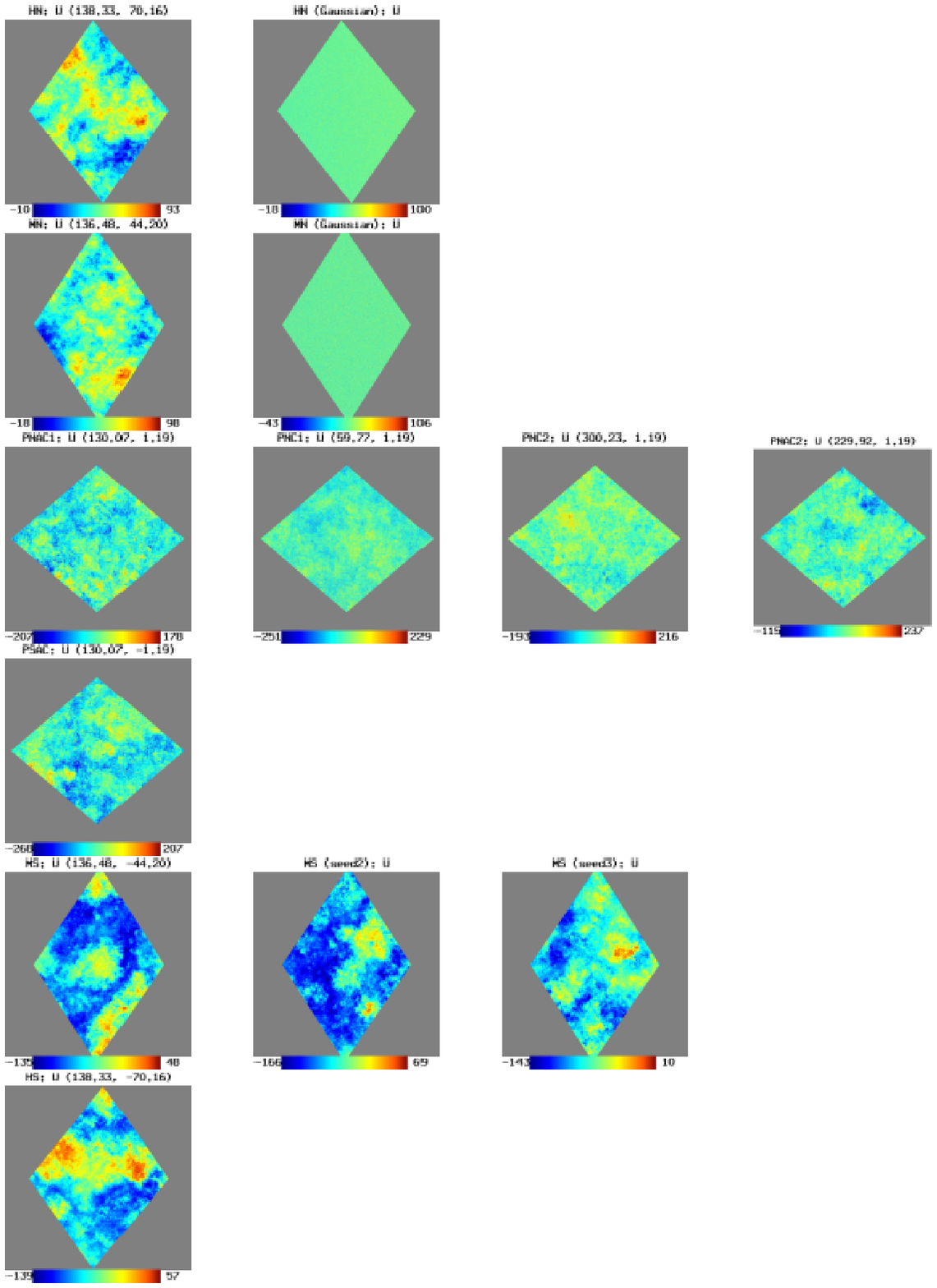}
\caption{As Fig.~\ref{imaps}, but for the simulated Stokes $U$ maps.}
\label{umaps}
\end{figure*}

\begin{figure*}[!htbp]
\centering
\includegraphics[width=18cm]{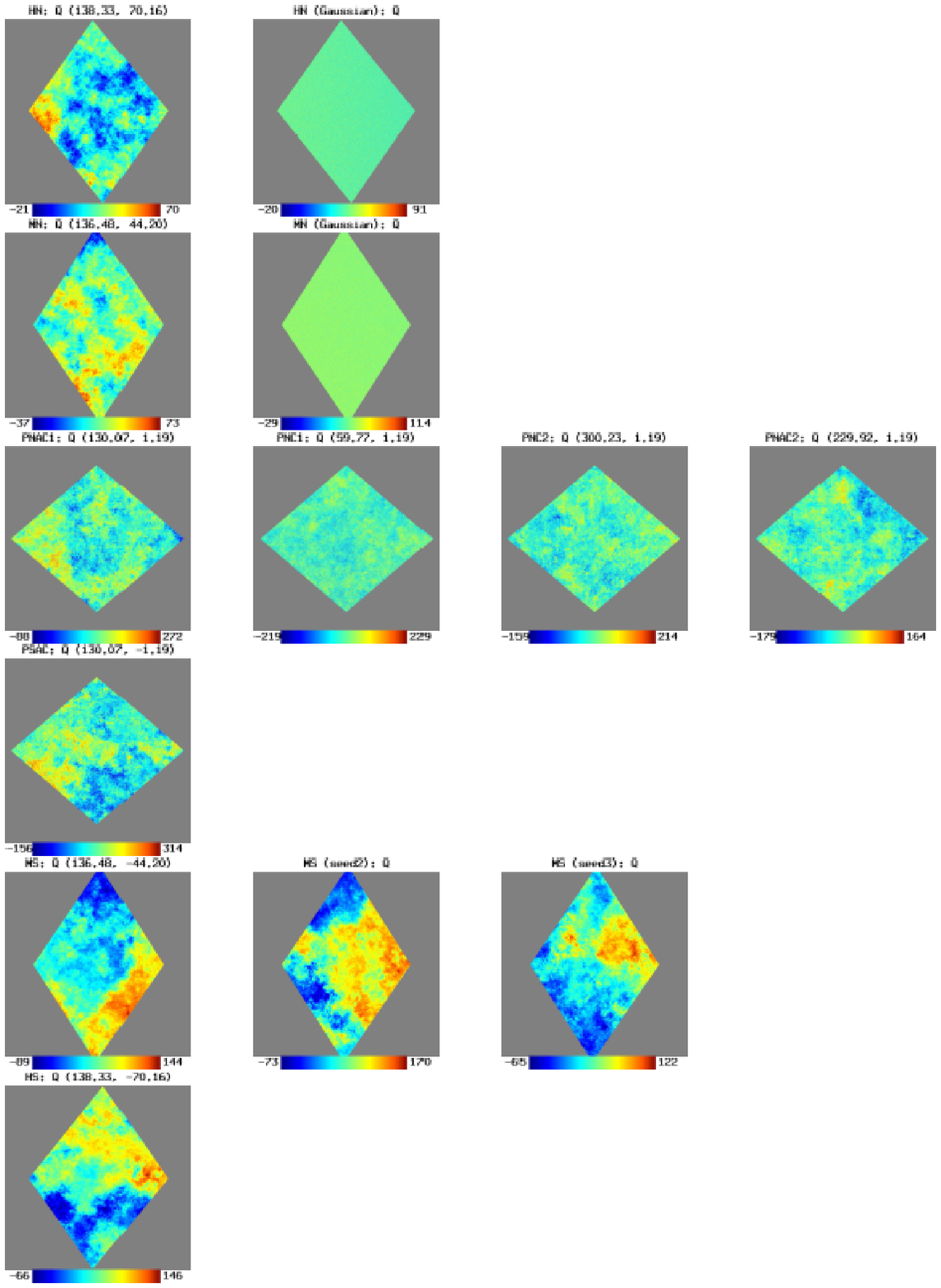}
\caption{As Fig.~\ref{imaps}, but for the simulated Stokes $Q$ maps.}
\label{qmaps}
\end{figure*}

\begin{figure*}[!htbp]
\centering
\includegraphics[width=18cm]{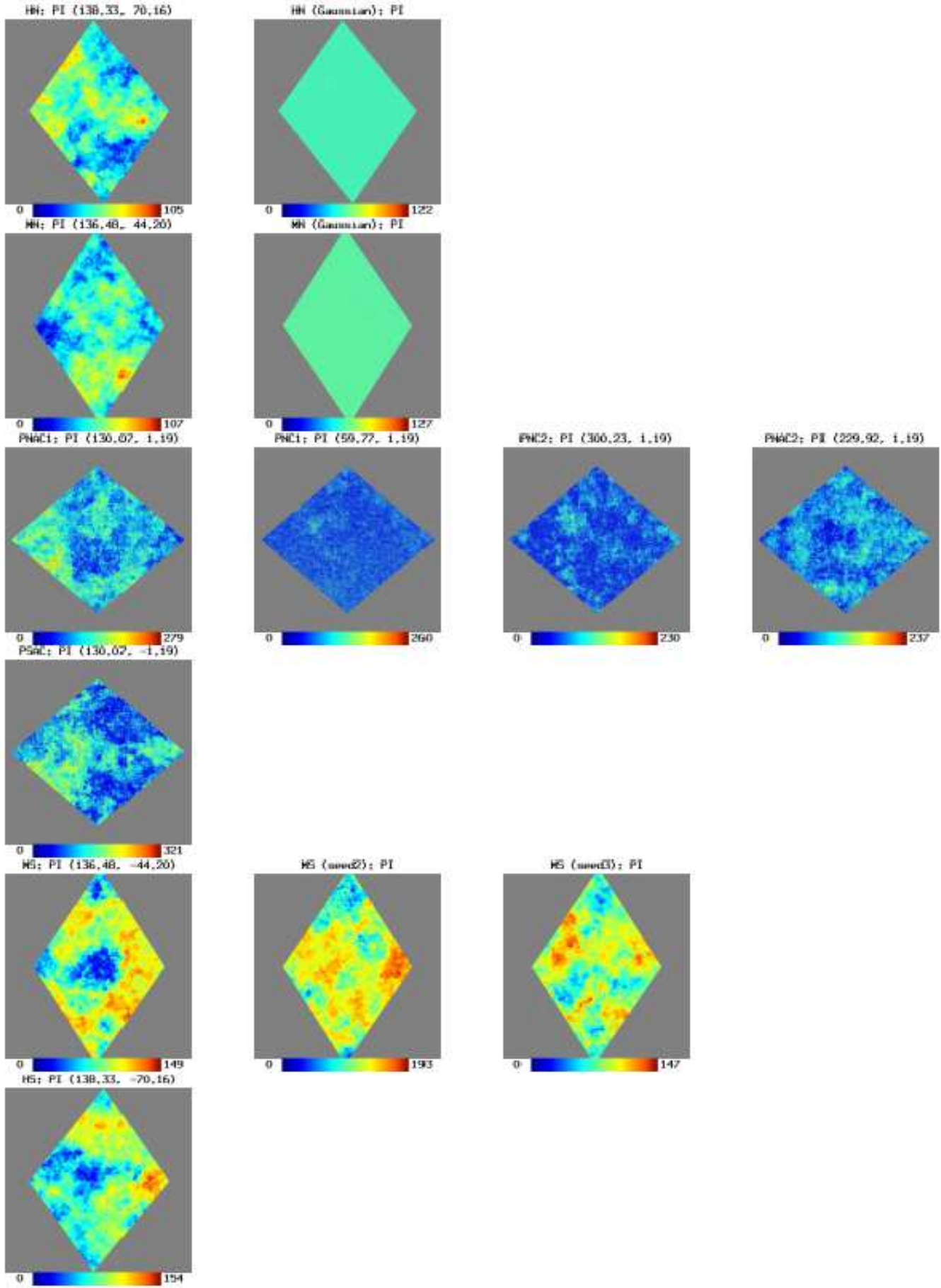}
\caption{As Fig.~\ref{imaps}, but for $PI$ maps calculated from the simulated 
$U$ and $Q$ maps.}
\label{pimaps}
\end{figure*}

\begin{figure*}[!htbp]
\centering
\includegraphics[width=18cm]{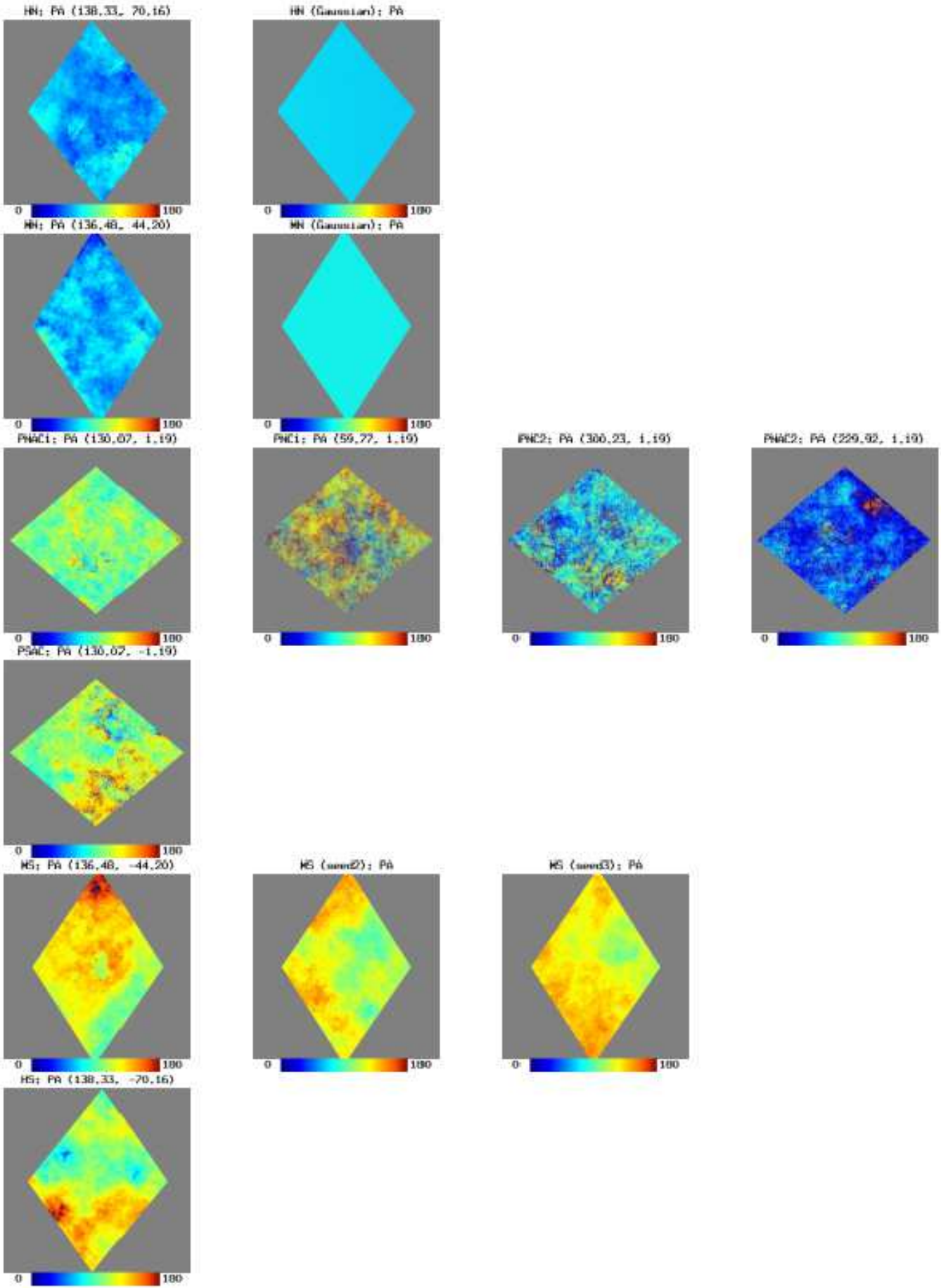}
\caption{As Fig.~\ref{imaps}, but for the polarization angles $PA$ calculated 
from the simulated  $U$ and $Q$ maps. 
}
\label{pamaps}
\end{figure*}

We display the total intensity $I$ maps in Fig.~\ref{imaps}, Stokes $U$ maps 
in Fig.~\ref{umaps}, Stokes $Q$ maps in Fig.~\ref{qmaps}, polarized intensity 
$PI$ maps in Fig.~\ref{pimaps} and polarization angle $PA$ maps in 
Fig.~\ref{pamaps}.

\end{document}